\documentstyle[11pt,aaspp4]{article}

\def \kms {{\rm km~s^{-1}}}

\def \Msun {{{\rm M}_{\sun}}}

\def \DM {{{\rm DM}}}

\def \gas {{{\rm gas}}}
\def \H {{{\rm H}}}
\def \halo {{{\rm halo}}}
\def \Hubble {{{\rm Hubble}}}

\def \H {{{\rm H}}}

\def \Hm {{{\rm H}^-}}
\def \HH {{{\rm H}_2}}

\def \nH {{n_\H}}

\def \nHm {{n_\Hm}}

\def \nd {{\dot{n}}}

\def \DM {{{\rm DM}}}
\def \IGM {{{\rm IGM}}}
\def \gas {{{\rm gas}}}
\def \Hubble {{{\rm Hubble}}}
\def \Jeans {{{\rm Jeans}}}
\def \Bondi {{{\rm Bondi}}}
\def \halo {{{\rm halo}}}

\def \BDM {\begin{displaymath}}
\def \EDM {\end{displaymath}}
\def \BEQ {\begin{equation}}
\def \EEQ {\end{equation}}
\def \BEQA {\begin{eqnarray}}
\def \EEQA {\end{eqnarray}}
\def \NN {\nonumber}
\def \BL {\begin{list}}
\def \EL {\end{list}}
\def \BENUM {\begin{enumerate}}
\def \EENUM {\end{enumerate}}
\def \BITEM {\begin{itemize}}
\def \EITEM {\end{itemize}}
\def \BARR {\begin{array}}
\def \EARR {\end{array}}
\def \BFIG {\begin{figure}}
\def \EFIG {\end{figure}}

\lefthead{KEPNER, TRIPP, ABEL \& SPERGEL}
\righthead{Mini-halo Absorption Signatures}

\begin{document}

\title{
Absorption Line Signatures of Gas in \\ Mini Dark Matter Halos}


 \author{
   Jeremy~Kepner\altaffilmark{1,4},
   Todd~M.~Tripp\altaffilmark{1},
   Tom~Abel\altaffilmark{2,3}, and
   David~Spergel\altaffilmark{1}
 }
 \altaffiltext{1}{Princeton University Observatory, Peyton Hall, Ivy Lane,
 Princeton, NJ 08544--1001 \\
 (jvkepner/tripp/dns)@astro.princeton.edu}
 \altaffiltext{2}{Laboratory for Computational Astrophysics,
 NCSA, University of Illinois at Urbana-Champaign,
 405 Mathews Ave., Urbana, IL 61801}
 \altaffiltext{3}{Max Planck Institute fur Astrophysik,
 Karl-Schwarzschild Str. 1, 85768, Garching, Germany \\
 abel@mpa-garching.mpg.de}
 \altaffiltext{4}{Current address: MIT Lincoln Laboratory, Lexington, MA}

\begin{abstract}

  Recent observations and theoretical calculations suggest that some QSO
absorption line systems may be due to gas in small dark matter halos
with circular velocities on the order of $30~\kms$.  \cite{Kepner97} (1997)
have shown that gas in these ``mini-halos'' can readily be in a
multi-phase state.  Additional observational evidence suggests that, in
general, many absorption line systems may also be multi-phase in nature. 
Thus, computing the absorption lines of mini-halos, in addition to
providing signatures of small halos, is a natural way to explore
multi-phase behavior.  The state of gas in mini-halos is strongly
affected by the background UV radiation field.  To address this issue a
code was developed that includes many of the chemical and radiative
processes found in CLOUDY and also incorporates spherically symmetric
multi-wavelength radiative transfer of an isotropic field,
non-equilibrium chemistry, heating, cooling and self-consistent quasi
hydro-static equilibrium gas dynamics.  With this code detailed
simulations were conducted of gas in mini-halos using different types of
background spectra: power-law, power-law + HeII break, \cite{Haardt96} (1996)
and O star.  From these simulations the absorption line signatures of
the gas were computed and compared with a variety of observations: high
redshift metal lines, He lines and low redshift metal line systems. 
Based on these results the mini-halo model absorption line signatures
appear to be consistent with many current observations given a
sufficiently soft spectrum.  Thus, in any given instance it is difficult
to either rule in or rule out a mini-halo, and in most cases additional
data (e.g.  optical counterparts or the lack thereof) or contextual
information (e.g.  evidence of significant star formation, which would
disrupt gas in a mini-halo) is necessary to break this degeneracy. 
Finally, the mini-halo model is a useful tool for analyzing absorption
line data in a multi-phase context and should become even more
applicable as new space based observations become available. 

\end{abstract}



\section{Introduction}

  Quasar absorption lines due to metals are sensitive probes of physical
conditions and chemical abundances.  With current instrumentation they
can be detected from $z$ = 0 to $z >$ 4 and therefore can be used to
track the chemical and physical evolution of galaxies and the
intergalactic medium over most of the history of the universe.  Metals
are now routinely detected in all types of QSO absorption line systems
with HI column densities ranging from $\sim 10^{14}$ to $\gg 10^{20}$
cm$^{-2}$.  Typically these absorption systems are analyzed using
idealized single-phase photoionization models where the absorber is
treated as a constant density plane-parallel slab illuminated from one
side (e.g., \cite{Bergeron86} 1986; however, see \cite{Dona91} 1991;
\cite{Giroux94} 1994).  Recently these models have considered rather
detailed and realistic ionizing radiation fields including reprocessing
of the radiation as it propagates through intervening absorption systems
and intergalactic space (e.g., \cite{Giroux97} 1997; \cite{Khare97}
1997), but they still usually assume a single-phase medium.  However, by
analogy with the ISM in our own galaxy, it seems more probable that QSO
absorption lines arise in multi-phase media, and there is observational
evidence which suggests that this is indeed the case in some absorbers. 
For example, in some absorbers the single-phase slab models cannot
reproduce all of the observed metal column densities (\cite{Giroux94}
1994; \cite{Petitjean96} 1996; \cite{Tripp97} 1997; \cite{Church99}
1999).  It is also readily apparent in some absorbers that high
resolution profiles of low and high ionization stages do not have
identical component structure in velocity space; some components are
strong in high ion profiles and are weak or not detected in low ion
profiles (see, e.g., Figures 2-21 in \cite{Lu96} 1996).  This suggests
the presence of multiple absorbing phases similar to those observed in
the Milky Way ISM.  Not surprisingly, the densities are also not
constant within a given absorber component; by observing common
absorption lines toward closely-spaced images of a gravitationally
lensed QSO, \cite{Rauch97a} (1997a) has shown that the metal densities
vary on the scale of a few hundred parsecs or less (see his Figure 5). 
Therefore it now seems worthwhile to consider more sophisticated
photoionization models which allow for multiple absorbing phases and
changing densities along the line of sight. 

  In this paper the absorption line signatures are calculated for gas in
small dark matter halos (i.e., mini-halos) with small circular
velocities ($v_{c} \sim$ 30 km s$^{-1}$).  \cite{Kepner97} (1997) have
shown that gas in this type of potential can be a multi-phase absorber
with a core and an ionized envelope, and the character of the core
depends on the intensity of the UV background radiation.  As the
intensity of the UV background decreases, the core passes through three
stages characterized by the predominance of ionized, neutral, and
molecular hydrogen (see
Figures~\ref{fig:HI_den_evolution}--\ref{fig:T_evolution}).  The model
includes full radiative transfer, gas dynamics, and non-equilibrium
chemistry and produces physically self-consistent hydrostatic gas density
and temperature radial profiles.  Given the extragalactic UV background
as input, the model can track the properties of the mini-halos from $z
\gg$ 4 to $z$ = 0.  Since previous papers have shown the detailed
dependencies of the metal line ratios on the assumed shape of the UV
background (e.g., \cite{Giroux97} 1997; \cite{Songaila98} 1998), the
primary goal of this paper is to explore the effects of the two-phase
core-envelope structure on the metal ratios and to determine if
mini-halos have any distinctive absorption signatures. 

  Observationally, there is evidence that the mini-halo model is a
plausible model for some QSO absorbers.  \cite{Abel98} (1998) have
suggested that if density perturbations collapse to form mini-halos
before reionization, then due to their high densities, the mini-halos
will remain largely neutral when the UV background turns on resulting in
a population of objects with N(HI) $\gtrsim 10^{17}$ cm$^{-2}$ which can
explain the number density of Lyman limit (LL) absorbers observed at
high redshifts.  Likewise, the simulations of \cite{Bond97} (1997)
predict large numbers of mini-halos with $v_c \sim 30~\kms$.  If the
mini-halos form after reionization, then they will be substantially more
ionized.  This is the model considered in this paper -- the objects
begin fully ionized and subsequently develop self-shielded cores as the
background intensity decreases.  One objection to the mini-halo model is
that it cannot explain the complex component structure and velocity
spread usually observed in QSO heavy element absorption profiles. 
However, in the hierarchical model of galaxy formation, ensembles of
dwarf-like objects coalesce to form larger galaxies, and in this case
the individual components in the coalescing object may be well-described
as mini-halos.  \cite{Rauch96} (1996) have shown that the two-point
correlation function of high z \ion{C}{4} absorbers is consistent with
the hierarchical formation scenario.  In this scenario, the number of
mini-halos should decrease with redshift as they merge into larger
systems.  Nevertheless, some of the mini-halos may survive down to $z$ =
0, and \cite{Blitz98} (1998) have recently suggested that the more
distant high velocity clouds in the vicinity of the Milky Way are in
fact mini-halos which have not yet accreted onto the galaxy. 

  Some higher column density QSO absorbers may also be due to
mini-halo-like objects.  For example, \cite{Steidel97} (1997) have been
unable to identify the damped Ly$\alpha$ system at $z_{\rm abs}$ = 0.656
in the spectrum of 3C 336 despite an exhaustive galaxy redshift survey
and deep {\it Hubble Space Telescope (HST)} and ground-based IR imaging. 
They conclude that this absorber is probably due to a dwarf galaxy with
$L < 0.05 L^*$ very close to the QSO.  This damped system has N(HI)
$\approx 2 \times 10^{20}$ cm$^{-2}$, and a wide variety of metals are
detected in this absorber in the {\it HST} Faint Object Spectrograph and
ground-based spectra of the QSO obtained by \cite{Steidel97} (1997), but
unfortunately none of their spectra have adequate resolution to compare
the absorption line kinematics to the mini-halo model.  On a different
sight line, \cite{Rao98} (1998) have identified two low redshift damped
Ly$\alpha$ absorbers in the spectrum of QSO OI 363, and they note that
``none of the galaxies visible in the vicinity of the quasar is a
luminous gas-rich spiral with low impact parameter,'' again raising the
possibility that these high column density systems are due to dwarf-like
objects.  \cite{Kepner97} (1997) have shown that when a mini-halo
attains a self-shielded \ion{H}{1} core, the \ion{H}{1} column density
in the core can exceed $10^{20}$ cm$^{-2}$.  However, if the damped
absorbers are due to several clustered mini-halos which will eventually
coalesce, then the N(\ion{H}{1}) of the individual mini-halos may be
lower while the {\it total} \ion{H}{1} column (integrated along the line
of sight) is sufficient to produce a damped absorber. 

  Finally, as suggested by \cite{Rees86} (1986) and \cite{Miralda93}
(1993), it is possible that some of the Ly$\alpha$ clouds may be due to
mini-halos, and \cite{Mo94} (1994) have shown that the observed number
density of Ly$\alpha$ clouds at various redshifts can be reproduced by
the mini-halo model.  \cite{Rees88} (1988) points out that due to
merging of mini-halos in the hierarchical galaxy formation model, at low
redshifts surviving mini-halos will be less likely to be found in
regions of large-scale overdensity.  Some of the recent studies of the
relationship between Ly$\alpha$ clouds and galaxies have found
Ly$\alpha$ clouds apparently in galaxy voids (e.g., \cite{Morris93}
1993; \cite{Stocke95} 1995; \cite{Tripp98} 1998); these may be
mini-halos which have survived to low $z$ by virtue of their location in
regions of low galaxy density. 

After the mini-halo model was introduced, it was criticized because a
huge number of halos per unit redshift would be required to reproduce
the observed density of absorption lines since the mini-halos have small
spatial cross-sections.  Also, observations of double sight lines to QSO
pairs indicate that some Ly$\alpha$ absorbers have very large spatial
extents (e.g., \cite{Dins95} 1995,\cite{Dins97} 1997).  It now seems
clear that {\it all} of the absorbers cannot be attributed to
mini-halos.  However, recent hydrodynamic simulations of cosmological
structure growth suggest that a variety of phenomena cause QSO
absorption lines ranging from very large gaseous filaments to
mini-halo-like objects.  Furthermore, large numbers of mini-halos are
found within the large filamentary structures in simulations at high
redshift (e.g., \cite{Bond97} 1997) as well as simulations pushed to $z$
= 0 (\cite{Dave99} 1999).  Interestingly, recent \ion{H}{1} 21 cm
imaging has revealed this predicted type of structure at very low
redshift: \cite{Hoff99} (1999) have discovered three mini-halo-like
objects embedded in the much larger gas envelope which surrounds the Sm
galaxies NGC 4532 and DDO 137 in Virgo.  These objects have the expected
masses of mini-halos and show no traces of star formation in deep CCD
images in B and R. 

  Given these observational and theoretical motivations, we have
revisited the mini-halo model for QSO absorption lines.  The rest of
this paper is organized as follows.  \S 2 presents the basic physical
model behind the mini-halo.  \S 3 describes the code used to compute
properties of mini-halos.  \S 4 discusses the various input spectra and
presents comparisons of observations with the mini-halo absorption
signatures.  \S 5 discusses the results and \S 6 gives our conclusions. 

\section{Mini-halo Model}

  The simulations attempt to follow the evolution of the gas in a fixed
halo potential.  For these purposes, the dark matter halo is specified by
two parameters: the circular velocity $v_c$ and the virialization
redshift $z_v$, which can be translated into a halo radius $r_\halo$ and
halo mass $M_\halo$ by assuming that the overdensity at virialization
is $\delta = 18 \pi^2$ (\cite{Gunn72} 1972)
  \BEQ
       v_c^2 = \frac{G M_\halo}{r_\halo} ~ , ~~~ 
      \frac{4\pi}{3} r_\halo^3 \delta \rho_c(z) = M_\halo ,
  \EEQ 
where the mean density is given by the usual expressions for a
$\Omega=1$ CDM cosmological model: $\rho_c(z) = (1
+ z)^3 \rho_c^0$, $6 \pi G \rho_c t_\Hubble^2 = 1$, $t_\Hubble^0 = 2/3 H_0$.  

  The dark matter halo profile is taken from \cite{Burkert95} (1995) and is based
on fits of dwarf galaxy rotation curves,
  \BEQ
     \rho_\DM(r) = \frac{\rho_0}{(1 + x) (1 + x^2)} ~, ~~~  x = r/r_0
  \EEQ
which in turn can be related to the halo radius and mass by
  \BEQA
       r_\halo &=& 3.4 r_0 , \NN \\
       M_\halo &=& M_\DM(r_\halo) , \NN \\
       M_\DM(r)  &=& \int_0^{r} \rho_\DM(r) 4 \pi r^2 dr .
  \EEQA
While recent numerical work suggests that the halo density profiles of
large galaxies are proportional to $r^{-1}$ in the centers and $r^{-3}$
at the edges (\cite{Navarro96b} 1996), these profiles do not fit the
dwarf galaxy observations (\cite{Moore94} 1994; \cite{Flores94} 1994). 

  The ultraviolet background is able to heat the gas to a temperature of
roughly $10^4~\arcdeg$K.  In large halos, where $v_c > 50~\kms$, the gas
pressure is relatively unimportant and the gas content is determined by
the global value of $\Omega_b$: $M_\gas = \Omega_b M_\halo$ (assuming
$\Omega=1$).  However, for smaller halos collapsing out of a hot IGM the
gas pressure resists the collapse (\cite{Thoul96} 1996) and $M_\gas <
\Omega_b M_\halo$.  A simple estimate as to where this transition occurs
and how much gas should reside in the halo can be made using the Jeans
mass and Bondi accretion limits.  If the gas mass in the uncollapsed
halo is greater than the Jeans mass, then the gas should collapse of its
own accord.  This provides an upper limit to amount of gas in the halo
 \BEQ
      M_\gas = \Omega_b M_\halo ~, ~~~ \Omega_b M_\halo > M_\Jeans
 \EEQ
where 
$M_\Jeans = f \Omega_b \rho_c (2 \pi c_\IGM t_\Hubble)^3$, $f = 3/\pi \sqrt{2}$,
$c_\IGM^2 = 1.5 k_B T_\IGM / \mu$.  For $\Omega_b M_\halo < M_\Jeans$ 
an upper limit can be computed for $M_\gas$ by estimating the amount of mass
that could be accreted via Bondi accretion in a Hubble time.  Thus,
 \BEQ
      M_\gas = M_\Bondi ~, ~~~ \Omega_b M_\halo < M_J
 \EEQ
where $M_\Bondi = f t_\Hubble \dot{M}_\Bondi$, $\dot{M}_\Bondi = \pi G^2
M_\halo^2 \rho_c / c_\IGM^3$. Note, the $O(1)$ factor $f$ has been included
in $M_\Jeans$ and $M_\Bondi$ so that $M_\Jeans = M_\Bondi$ when
$\Omega_b M_\halo = M_\Jeans$.

\section{Simulating Mini-halos}

  Calculating the absorption signatures is significantly complicated by
the large effect a background UV radiation field can have on gas in
small halos (\cite{Dekel86} 1986; \cite{Efstat92} 1992; \cite{Quinn96}
1996; \cite{Kepner97} 1997).  To address this issue a code was developed
that includes many of the chemical and radiative processes found in
CLOUDY (\cite{Ferland98} 1998) and also incorporates multi-wavelength
radiative transfer, non-equilibrium chemistry and gas dynamics.  The
full details of the code are given in \cite{Kepner97} (1997) and are
briefly summarized here. 

  The code computes the quasi-hydrostatic equilibrium states of gas in
spherically symmetric dark matter halos (roughly corresponding to dwarf
galaxies) as a function of the amplitude of the background UV field. 
The code integrates the full equations of radiative transfer, heating,
cooling and non-equilibrium chemistry for nine species of H and He
including H$_2$, as well as all the ionization states of the metals C,
O, Mg and Si.  These metals were chosen because they are commonly
observed in absorption line systems.  The density and temperature
profiles are evolved through an iterative procedure.  The initial gas
density profile is specified by hydrostatic equilibrium and by our
assumption that the gas is in thermal equilibrium with the background
radiation field.  At each redshift a new equilibrium temperature profile
is computed for current value of the background radiation field, which
evolves with redshift.  For a given temperature profile, DM potential
and total gas mass, it is then a simple matter to compute the density
profile necessary to maintain hydrostatic equilibrium.

  The important role of the detailed chemistry of primordial gas (in
particular the formation of $\HH$) has been known and studied since it
was first proposed as a mechanism for the formation of globular clusters
(\cite{Peebles68} 1968).  The potential number of reactions in this
simple mixture of H and He is enormous (\cite{Janev87} 1987). 
\cite{Abel97} (1997) have selected a subset of these reactions to model
the behavior of primordial gas for low densities ($n < 10^4$ cm$^{-3}$)
over a range of temperatures ($1^\circ$K $< T < 10^8~^\circ$K).  Among
the processes included in this model are the photo-attachment of neutral
hydrogen, the formation of molecular hydrogen via H$^-$, charge exchange
between H$_2$ and H$^+$, electron detachment of H$^-$ by neutral
hydrogen, dissociative recombination of H2 with slow electrons,
photodissociation of H$_2^+$, and photodissociation of H$_2$.  In
addition, these species have been supplemented with the appropriate
chemical and radiative processes for four commonly detected metals: C,
O, Mg, and Si.  For the metals, this required obtaining additional
collisional ionization rate coefficients (\cite{Voronov97} 1997),
radiative recombination rates (\cite{Verner96a} 1996a), dielectronic
recombination rates (\cite{Aldrovandi73} 1973; \cite{Shull82} 1982;
\cite{Arnaud85} 1985; \cite{Nussbaumer83} 1983), charge exchange rates
(\cite{Kingdon96} 1996), and photoionization cross sections
(\cite{Verner96b} 1996b; \cite{Verner95} 1995). 

  Fully 3D radiative transfer requires estimating the contribution to
the flux at every point from every other point along all paths for each
wavelength.  At the minimum this is a 6D problem.  However, in most
instances symmetries can be introduced which result in a more tractable
situation.  The simplest situation occurs when the gas can be assumed to
be optically thin throughout.  This approximation is sufficient in the
majority of cosmological situations.  The next simplest geometry is that
of a slab (or a sphere under the assumption of a radially perpendicular
radiation field), which leaves an intrinsically 2D problem.  Although
this approach may not be a bad approximation for a sphere in an
isotropic radiation field, this code accounts for all the different
paths that penetrate a given spherical shell, which leaves an inherently
3D problem.  Taking into account the different paths effectively
``softens'' the optical depth, smoothing out transitions from optically
thin to optically thick regimes. 

  Perhaps the most important aspect of the model is the balance between
the heating and cooling processes.  This balance is what allows the
establishment of a quasi-static temperature profile for a specific
radiative flux.  If the balance between the heating and cooling is not
established, then the hydrostatic equilibrium solution to the gas
profile will evolve too rapidly.  Fortunately, this situation only comes
about when the gas in the halo becomes dense and a large amount of H$_2$
is formed.  This point presumably marks the onset of star formation,
which would dramatically alter the situation, and so the calculation is
halted when H$_2$ cooling dominates.  The temperature profile is evolved
via the heating and cooling functions found in \cite{Anninos97} (1997), which
includes photoionization heating and cooling due to collisional
excitation, collisional ionization, recombination, molecular hydrogen,
bremsstrahlung and Compton cooling. 

  The microphysical processes couple to the larger scale density profile
primarily through radiative heating, which sets the temperature profile. 
The rate of radiative heating (primarily due to HI, HeI and HeII) is in
turn strongly dependent on the column densities of each species, which
is set by the temperature.  Thus, the resulting system is described by
differential equations on the small scale with integral constraints on
the large scale.  The difficulty of solving such a system is the large
variety of time scales involved.  Solving the entire set simultaneously
is prohibitive.  The approach taken here has been to use code modules
which solve for each of the processes independently.  Iterating between
the modules then provides an adequate approximation to the true solution
(see \cite{Kepner97} 1997 for a more complete discussion). 

\begin{figure}
\plotone{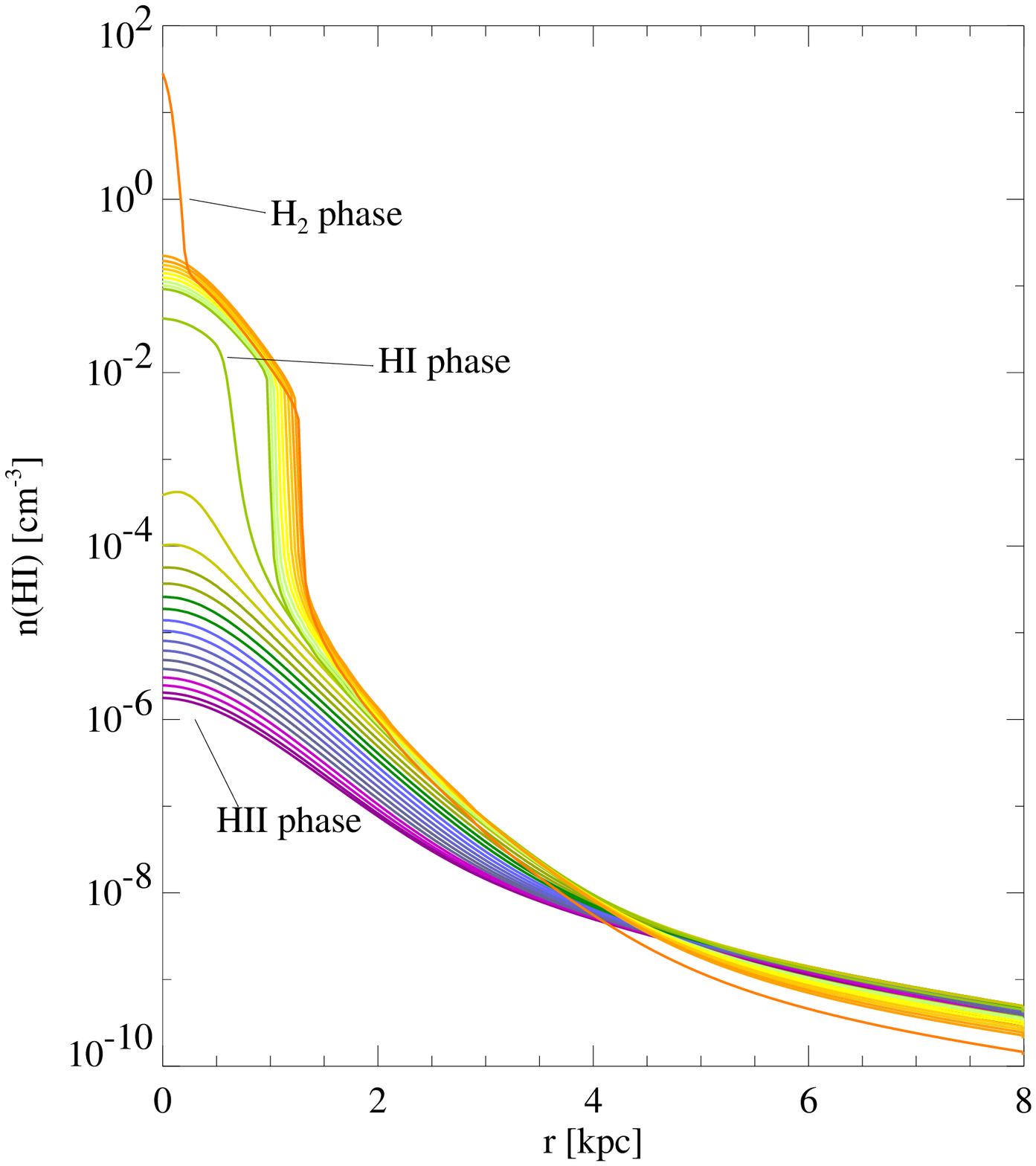}
\caption{
\label{fig:HI_den_evolution}
  Evolution of the HI number density profile of a typical object ($v_c
\approx 30~\kms$, $z_v = 3$, $M_\halo = 2.8\times 10^9~\Msun$, $M_\gas =
1.5\times10^7~\Msun$) from higher flux to lower flux.  As the
radiation flux decreases the behavior of the core is characterized by three
phases: HII, HI and H$_2$.  In the HII phase the density profile is
relatively smooth.  In the HI phase a core with a radius of $\sim$1 kpc
develops with a density of $\sim 0.1~cm^{-3}$.  In the H$_2$ phase rapid
cooling causes a dramatic increase in the density in the core to $\sim
10~cm^{-3}$. 
}
\end{figure}

\begin{figure}
\plotone{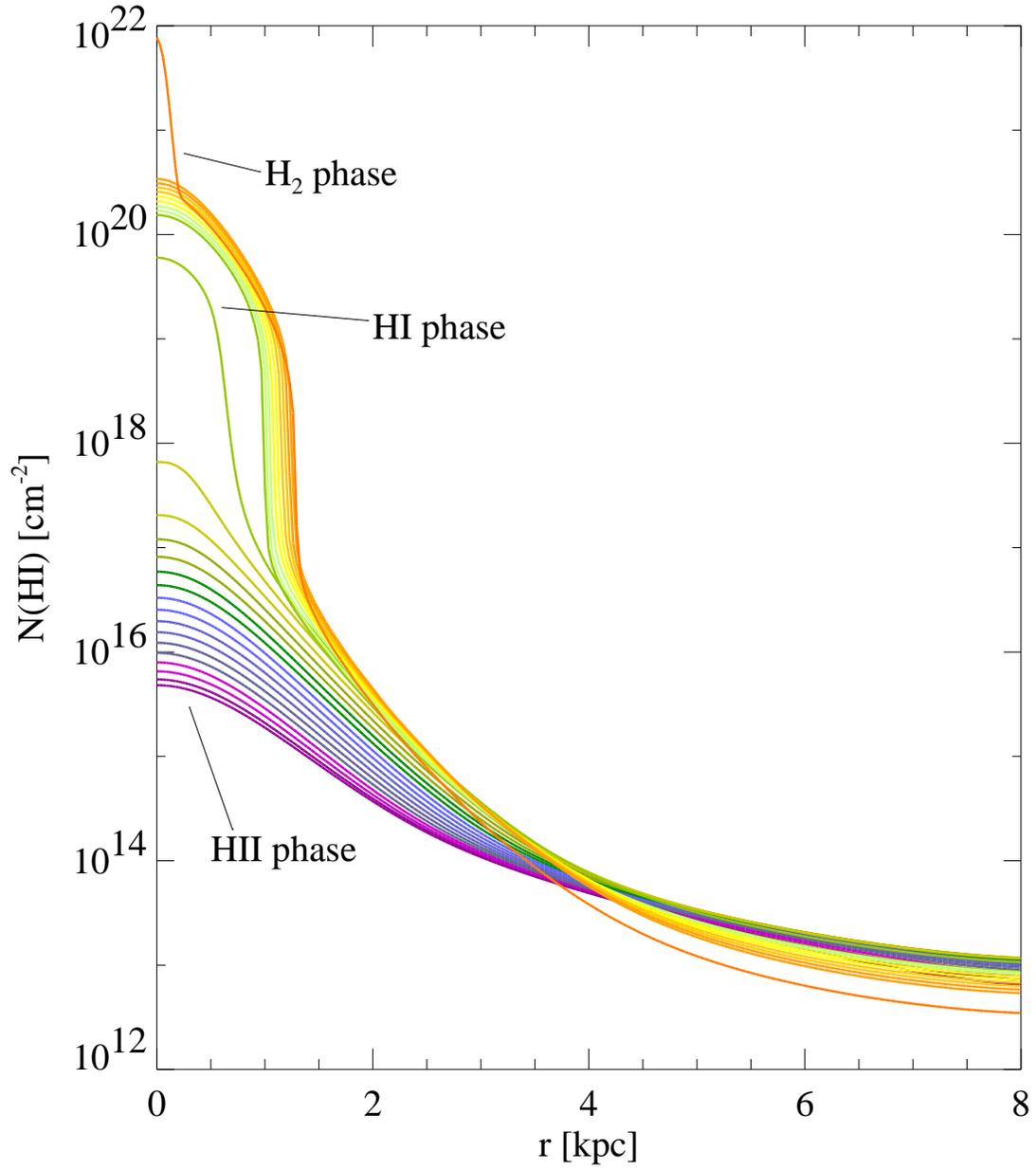}
\caption{
\label{fig:HI_col_evolution}
  Evolution of the HI column density profile for the object shown
in Figure~\ref{fig:HI_den_evolution}.
}
\end{figure}

\begin{figure}
\plotone{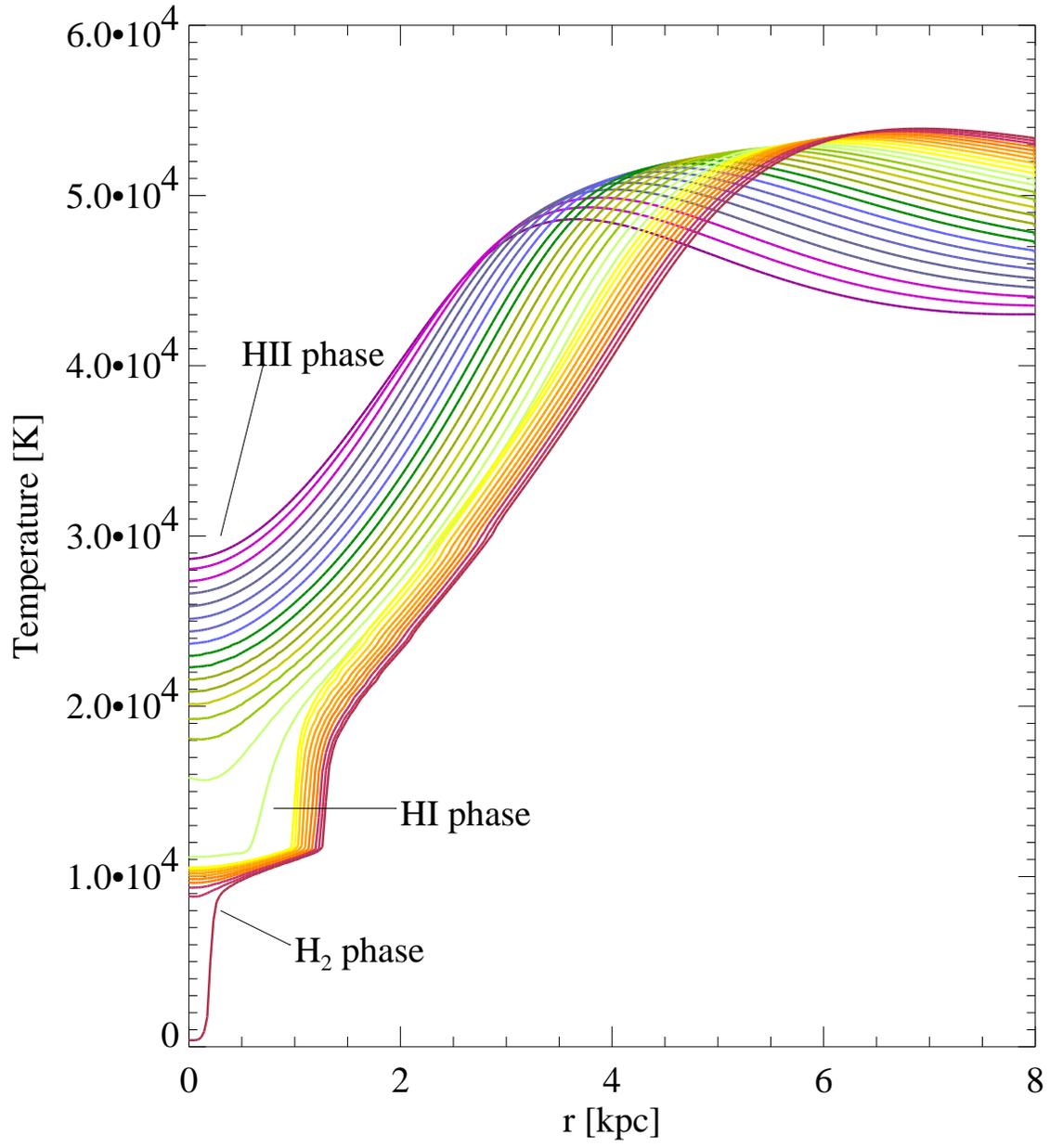}
\caption{
\label{fig:T_evolution}
  Evolution of the temperature profile for the object shown
in Figure~\ref{fig:HI_den_evolution}.
}
\end{figure}

\begin{figure}
\plotone{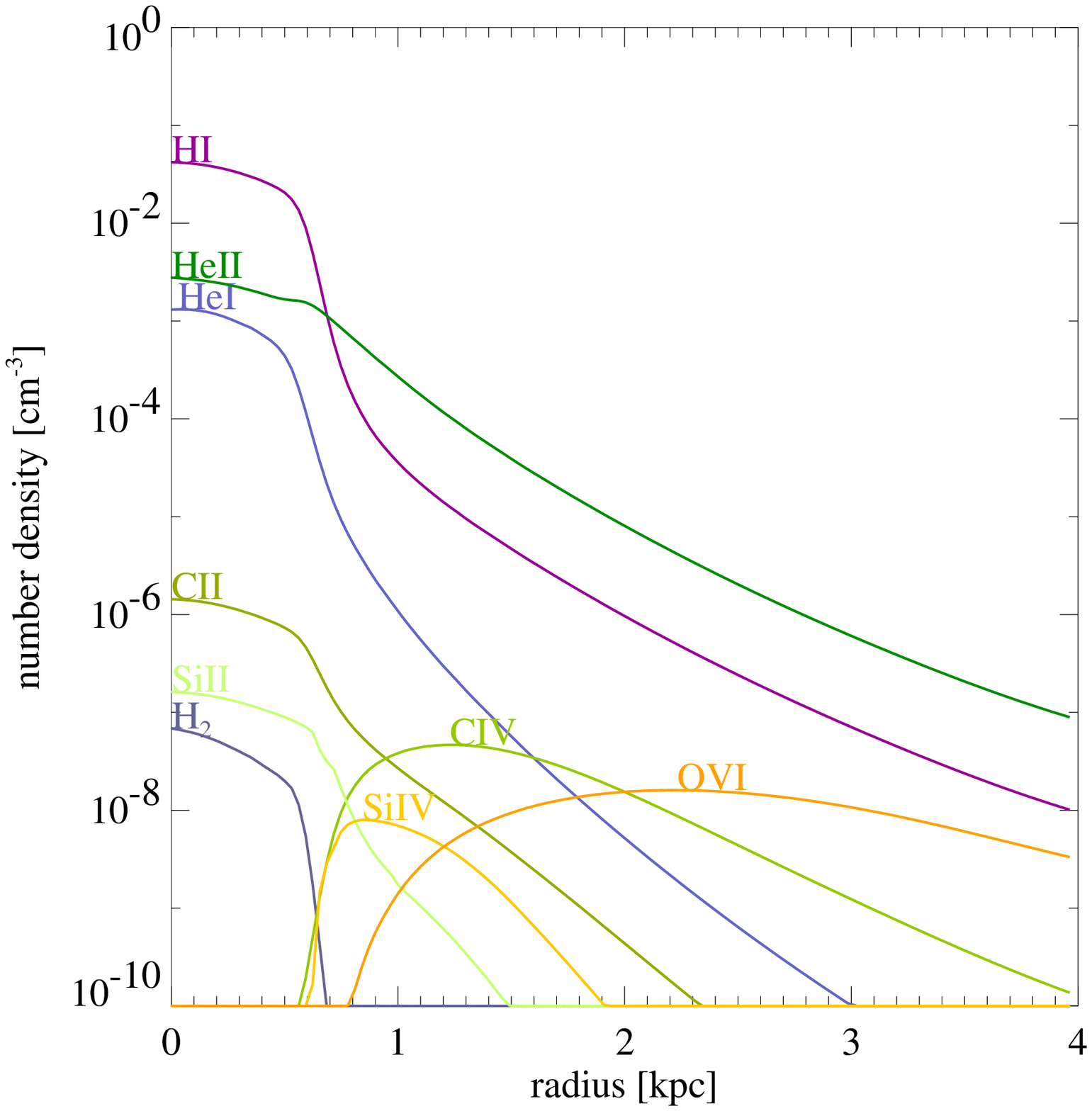}
\caption{
\label{fig:HI_phase_den}
  Number density profiles of HI, H$_2$, HeI, HeII, CII, CIV, SiII, SiIV
and OVI during the HI phase for the object shown in
Figure~\ref{fig:HI_den_evolution}.
}
\end{figure}

\begin{figure}
\plotone{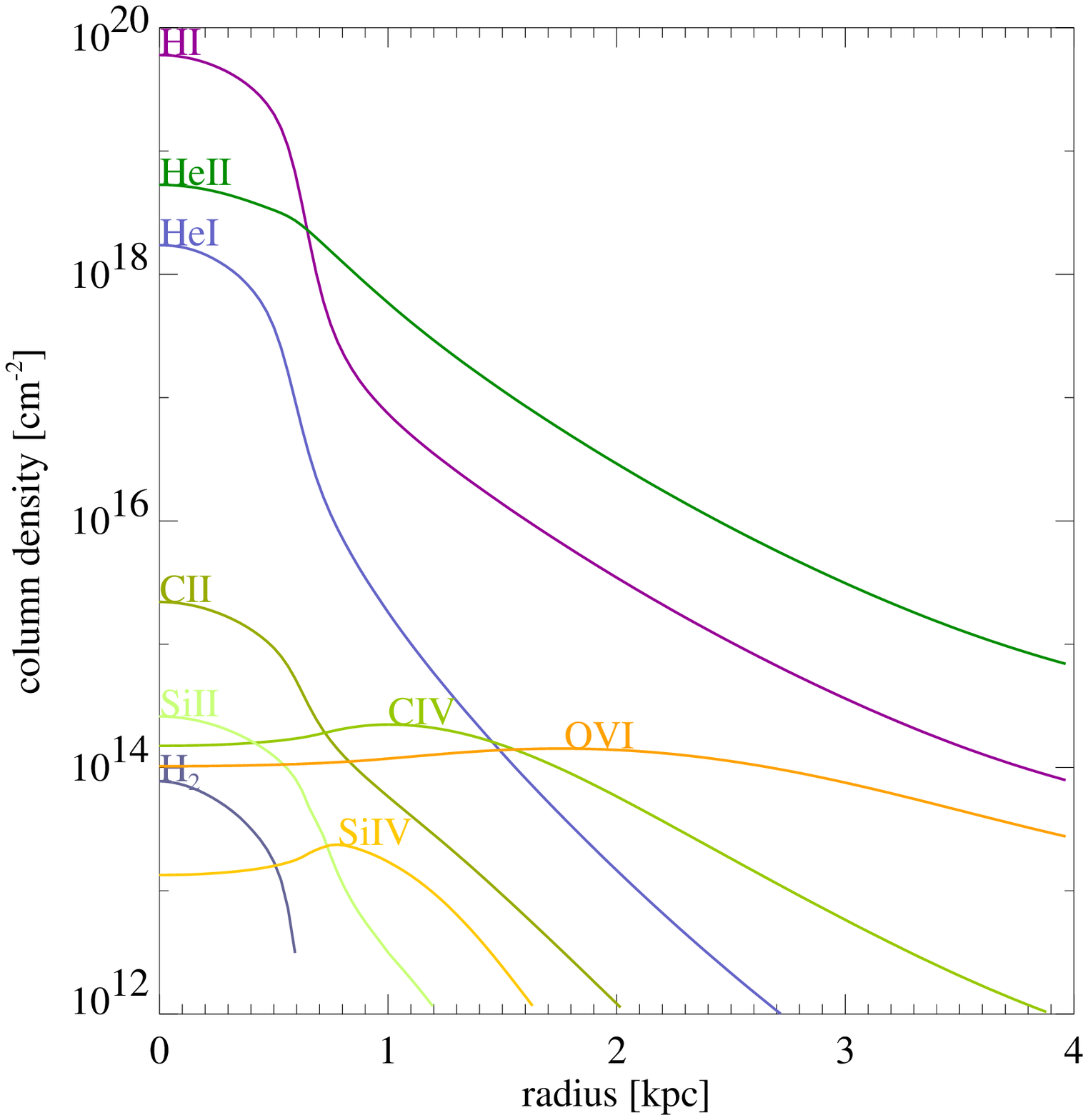}
\caption{
\label{fig:HI_phase_col}
  Column densities profiles of HI, H$_2$, HeI, HeII, CII, CIV, SiII, SiIV
and OVI during the HI phase for the object shown in
Figure~\ref{fig:HI_den_evolution}..
}
\end{figure}

  For a given input spectrum (e.g.  power law with $\alpha = -1.5$), as
the amplitude of the UV background is decreased the gas in the core of
the dwarf goes through three stages characterized by the predominance of
ionized (HII), neutral (HI) and molecular (H$_{2}$) hydrogen.  The last
stage (H$_{2}$) marks the onset of runaway cooling and presumably
star-formation.  Figures~\ref{fig:HI_den_evolution},
\ref{fig:HI_col_evolution}, and \ref{fig:T_evolution} show this
evolution as illustrated by the HI density, HI column density and gas
temperature profiles.  Figures~\ref{fig:HI_phase_den} and
\ref{fig:HI_phase_col} show the number density profiles and computed
column density profiles of several species for a mini-halo in the
neutral (HI) phase. 

  Although these calculations include C, O, Mg and Si for computing the
strengths of absorption lines, metals are ignored in the cooling of the
gas.  Metals can have a number of important effects on the chemistry
and dynamics of gas clouds: dust grains will absorb ionizing radiation
and serve as formation sites for molecular hydrogen; atomic lines of C
and other heavy elements can be important coolants.  Are these processes 
important in dwarf galaxies
at high redshift? Observations of QSO Lyman forest clouds suggest that
the metal abundances in meta-galactic gas is $Z \sim 0.001 - 0.01$
times the solar value at $z \sim 3$ (\cite{Songaila96} 1996).  At these
abundances, heavy element cooling is unimportant (\cite{Bohringer89} 1989). 
An upper limit on the mass in dust grains can be obtained by assuming
that most of the carbon at high redshift is incorporated into dust
grains.  If the size distribution of the grains is similar to the local
ISM, then this suggests a cross-section per hydrogen atom of
$\sigma_{\rm dust}(1000\AA) \approx Z~2 \times 10^{-21}$cm$^2$
(\cite{Draine96} 1996).  In the mini-halo model, the maximum column density
occurs when the cloud is most centrally condensed, which occurs just
before the onset of $\HH$ formation and is roughly N(HI) $\sim
10^{21}$cm$^{-2}$ (see Figure~\ref{fig:HI_den_evolution}).  Thus the
maximum optical depth at these wavelengths is approximately $\tau_{\rm
dust} \sim Z \sim 0.01$.  The contribution of dust to H$_2$ formation
in our galaxy can be approximated by a $\nd_\HH \approx R n \nH$, where
$R = 6 \times 10^{-18} T^{1/2}~{\rm cm}^3~{\rm s}^{-1}$
(\cite{Draine96} 1996).  If we scale $R$ by the metallicity, then the dust
term will be negligible in comparison to the other terms contributing
to $\HH$ formation whenever $\nHm > 10^{-9}$cm$^{-3}$, which is nearly
always the case in the neutral H core. 

\begin{figure}
\plotone{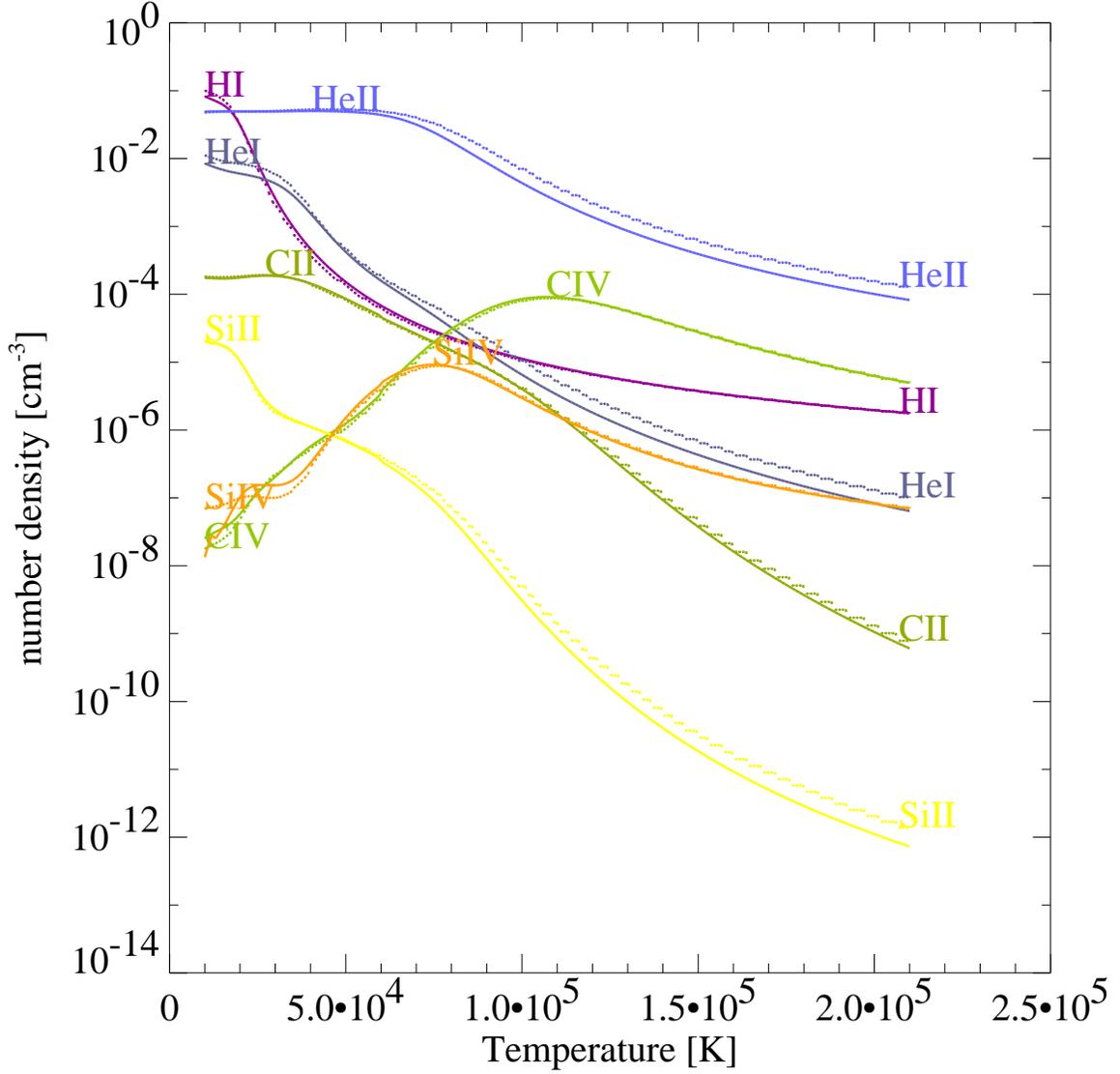}
\caption{
\label{fig:cloudy_comparison}
  Number density vs.  gas temperature of HI, HeI, HeII, CII, CIV, SiII
and SiIV for an optically thin slab of constant density illuminated by a
power-law ($\alpha = -1$) spectrum.  Solid lines are results of our
code.  Dashed lines are the results from a similar calculation performed
with CLOUDY. 
}
\end{figure}

  As a test of the accuracy of the chemistry and radiative parts of the
code, an optically thin static gaseous halo illuminated by simple
power-law spectrum was tested.  Figure~\ref{fig:cloudy_comparison}
compares the number densities of various species computed with the
mini-halo code with the results of a similar calculation performed with
CLOUDY.  The two codes are in good agreement. 

\section{Comparison with Observations}

  Having laid the groundwork for the mini-halo model, it is possible to
proceed with several observational comparisons.  Heavy element
absorption lines in the higher column density QSO absorbers have been
measured and studied intensively for some time.  With the advent of the
echelle spectrograph on the Keck telescope (\cite{Vogt94} 1994), it has
become routine to also detect metals in high redshift Ly$\alpha$ clouds
(\cite{Tytler95} 1995; \cite{Cowie95} 1995).  For example,
\cite{Songaila96} (1996) have reported that CIV is detected in 75\% of
Ly$\alpha$ clouds with N(HI)~$\gtrsim 3 \times 10^{14}$cm$^{-2}$. 

  Based on single-phase photoionization models constructed with CLOUDY,
\cite{Giroux97} (1997) and \cite{Songaila98} (1998) have shown that in
absorbers with N(HI) $\lesssim 10^{18}$ cm$^{-2}$, the observed SiIV/CIV
column density ratios at high redshift require an overabundance of Si
relative to C by factors of 2-3 and/or an ionizing spectrum which is
softer than expected based on detailed calculations of the background
due to QSOs and AGNs.  This can be achieved by a strong contribution to
the ionizing spectrum from local hot stars (\cite{Giroux97} 1997) or by
putting a large break in the ionizing spectrum at the HeII edge
(\cite{Songaila98} 1998).  In addition, \cite{Songaila96} (1996) and
\cite{Songaila98} (1998) have suggested that the SiIV/CIV ratio
decreases rapidly at $z \sim$ 3 (Note: other observations do not show
this change, e.g., \cite{Boksenberg97} 1997), which can be interpreted
as evidence that the ionizing spectrum changes abruptly at this redshift
perhaps due to the completion of intergalactic He reionization (see also
\cite{Heap99} 1999).

  It has also become possible to detect helium absorption in the spectra
of QSOs.  Absorption due to HeII has now been detected in the
spectra of several high $z$ QSOs with the {\it HST} 
(\cite{Jakobsen94} 1994; \cite{Tytler95} 1995; \cite{Hogan97} 1997; 
\cite{Reimers97} 1997; \cite{Heap99} 1999) as well as the Hopkins Ultraviolet 
Telescope (\cite{Davidsen96} 1996)).  In some cases voids in the HI Ly$\alpha$ 
forest closely match voids in the HeII absorption (\cite{Reimers97} 1997; 
\cite{Heap99} 1999), which suggests that a substantial portion of the He 
absorption is due to unresolved discrete HeII lines corresponding to the 
HI Ly$\alpha$ forest.  If many of the Ly$\alpha$ clouds are due to mini-halos, 
then the He absorption lines due to mini-halos should be consistent with the
observed He absorption.  Also, HeI absorption lines have been
reported in some Lyman limit absorbers (\cite{Reimers93} 1993).

  By stacking the spectra of low redshift Ly$\alpha$ clouds in the rest
frame, \cite{Barlow98} (1998) have shown that low $z$ clouds also
contain metals with apparently higher metallicities than their high
redshift counterparts (however, see \cite{Shull98} 1998 for a
counterexample of a rather low metallicity absorber at low redshift). 
Therefore observations of low $z$ QSOs with the Space Telescope Imaging
Spectrograph (STIS) should enable detailed studies of abundances and
physical conditions in the low redshift clouds whose optical
counterparts are more readily observable. 

  In this section these observations are compared to the predicted
properties of the mini-halo model.  \S 4.1 summarizes the various input
spectra assumed for the model.  The resulting metal line ratios are
presented in \S 4.2.  The observed high redshift metal-line ratios are
compared with the mini-halo metal line ratios in \S 4.3.  In \S 4.4 a
similar comparison is made with He lines.  Finally, as an example of
the kind of comparisons that will be possible in the future, the model
is applied to a low redshift OVI absorber in \S 4.5.  For the
calculations in this section the metallicity is fixed at 1/10 solar. 
The metal line ratios are independent of metallicity as long as the
metallicity is low enough so that radiative cooling by heavy elements
is not important, in which case the metal column densities directly
scale with N(HI) and metallicity. 

\subsection{Input Ionizing Spectra}

\begin{figure}
\plotone{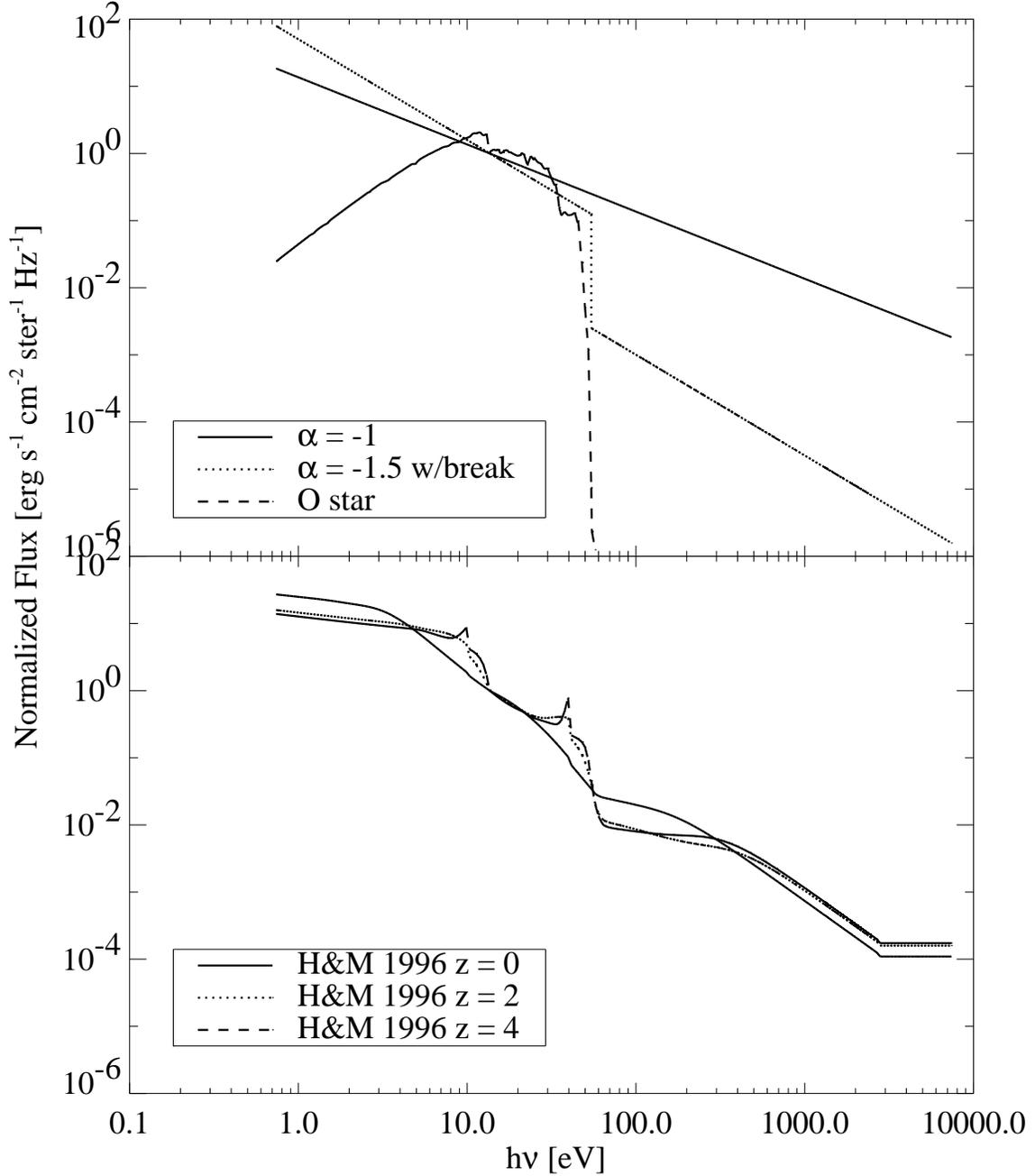}
\caption{
\label{fig:input_spectrum}
  Shape of the six different types of background spectra used to heat
gas in mini-halos.  The spectra have been normalized so that
$J_\nu$(h$\nu$ = 13.6 eV) = 1.0.  The top panel shows an $\alpha = -1$
power-law spectrum (solid line), an $\alpha = -1.5$ power-law spectrum
with a factor of 50 drop at 54.4 eV (dotted line), and an O star
spectrum (dashed line).  The bottom panel shows three spectra computed
by Haardt \& Madau 1996 at: $z = 0$ (solid line), $z = 2$ (dotted line)
and $z = 4$ (dashed line). 
}
\end{figure}

  Several spectral shapes are used as inputs: (1) a power-law with
spectral index $\alpha = -1$ (i.e., $f_{\nu} \propto \nu ^{-1}$), (2) a
power-law with $\alpha = -1.5$ and a factor of 50 break at the HeII
edge, (3) an O star with $T_{\rm eff}$ = 50,000$\arcdeg$K from the
ATLAS9 models computed and distributed by R.L.  Kurucz, and (4-6) the UV
background due to QSOs and AGNs calculated by \cite{Haardt96} (1996) at
redshifts z = 0, 2, and 4.  The power-law and power-law + HeII break
spectra are standard approximations of the UV background due to QSOs and
AGNs but neglect much of the radiation reprocessing which occurs in
intervening gas; the detailed effects of intervening absorption and
re-emission are included in the \cite{Haardt96} (1996) models.  The O star
spectrum is not particularly realistic for a real QSO absorber but is
useful for illustrating the differences between a mini-halo ionized by
UV radiation predominantly from local hot stars and a mini-halo
photoionized by the background radiation from distant quasars.  The
shapes of these spectra are shown in Figure~\ref{fig:input_spectrum}. 
The spectra have been normalized so that $J_\nu$(h$\nu$ = 13.6 eV) = 1.0. 
The mini-halo model evolves the spectra by decreasing its overall
amplitude, beginning high and slowly dropping off.  This is meant to
crudely approximate the evolution of the amplitude of the radiation
field with redshift.  As the amplitude of the radiation field drops, the
gas in the core of the mini-halo enters the HI phase, while the outer
parts are still ionized.  The various column densities computed from the
mini-halo that are shown in the subsequent sections are done so during
the HI phase, which provides the best example of a multi-phase absorber. 

\subsection{Mini-Halo Metal Line Ratios}

\begin{figure}
\plotone{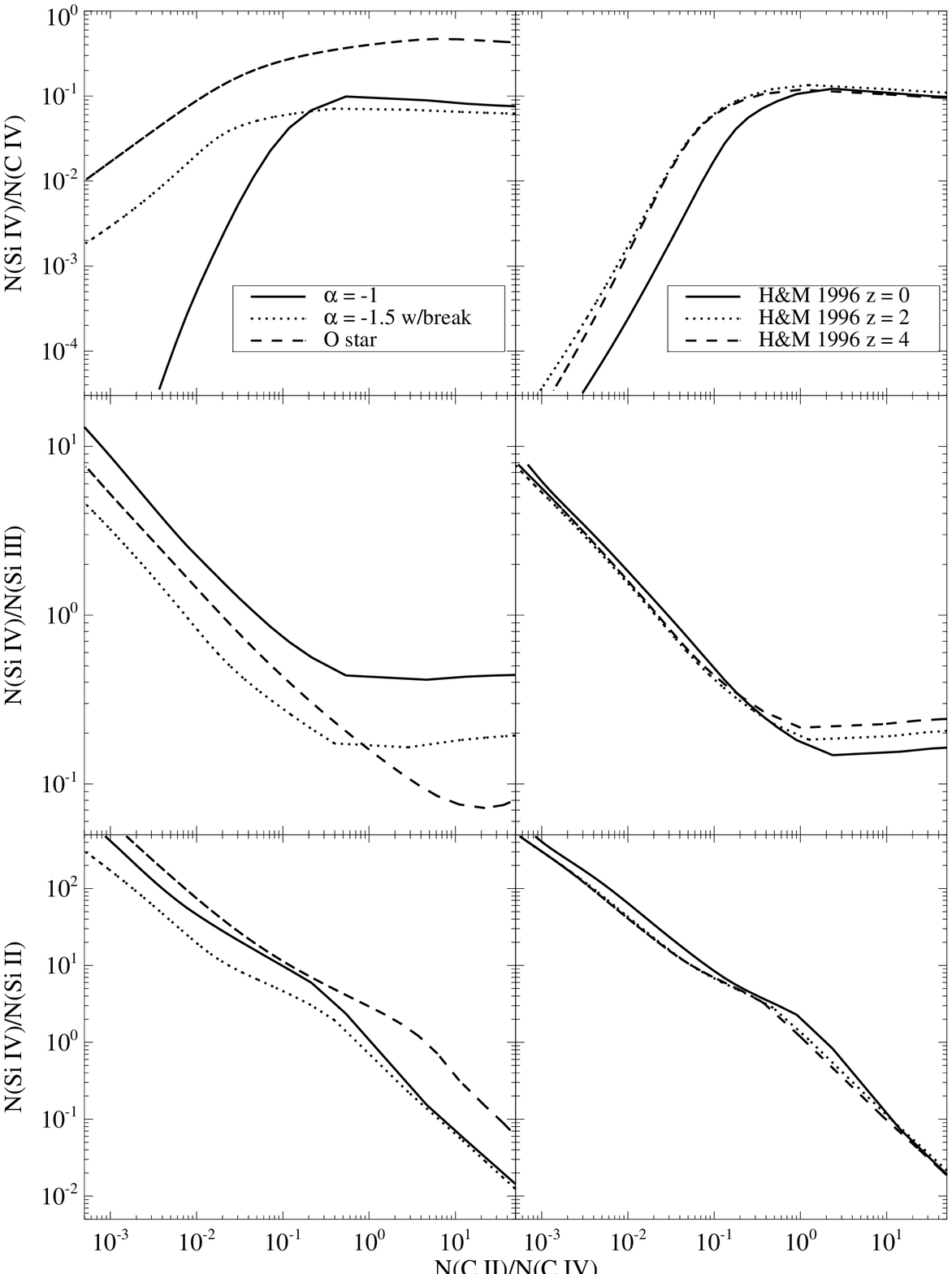}
\caption{
\label{fig:metal_ratios_a}
  The N(SiIV)/N(CIV),
N(SiIV)/N(SiIII) and N(SiIV)/N(SiII)
column density ratios vs.  N(CII)/N(CIV) as predicted
by the mini-halo model for the spectra shown in
Figure~\ref{fig:input_spectrum}.  The left panels plot the ratios
corresponding to the $\alpha = -1$ (solid line), $\alpha = -1.5$ w/break
(dotted line) and O star spectra (dashed line).  The right panels plot
the ratios from using the spectra generated by Haardt \& Madau 1996 at
three redshifts: z = 0 (solid line), z = 2 (dotted line) and z = 4
(dashed line). 
}
\end{figure}

\begin{figure}
\plotone{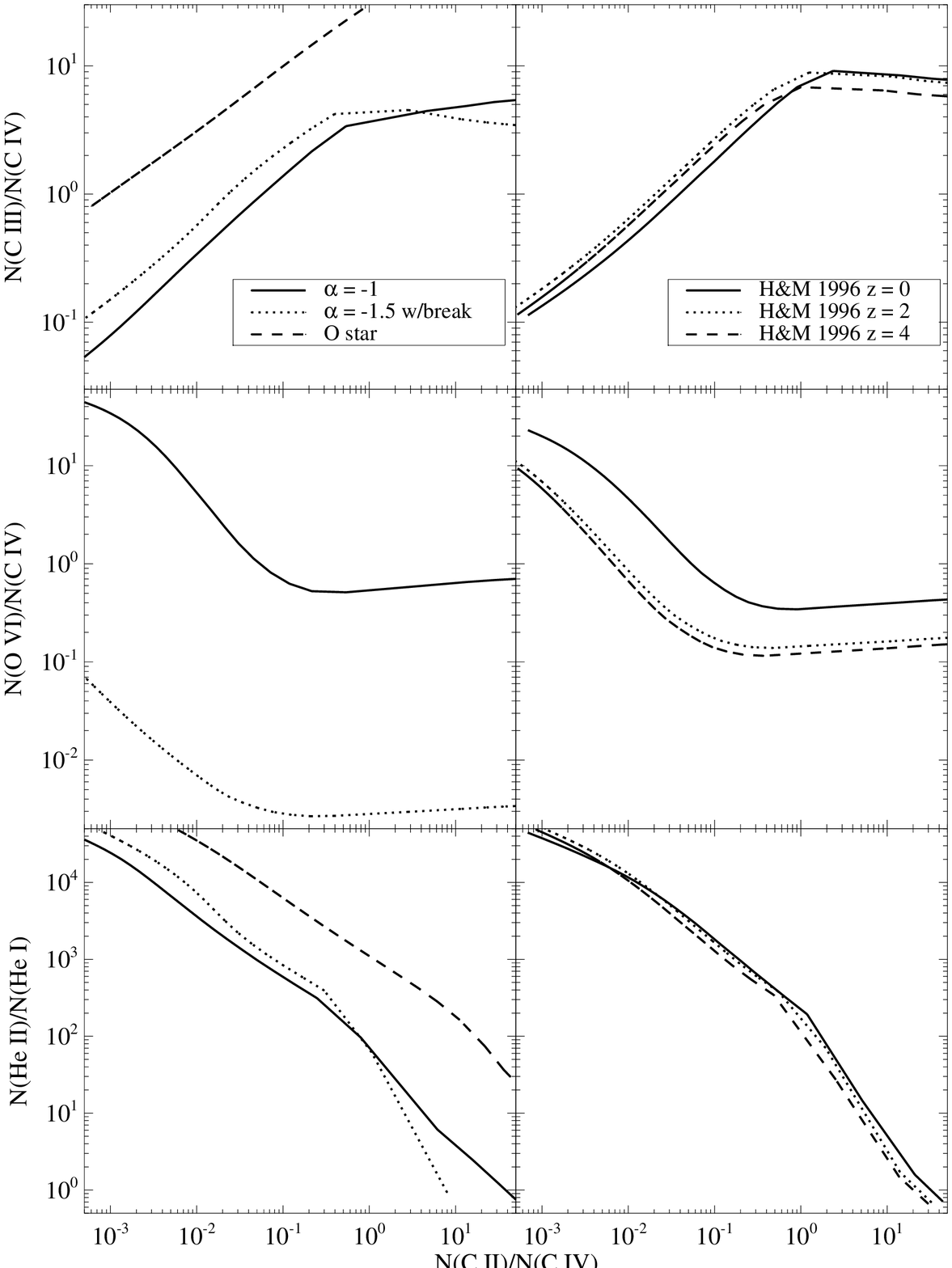}
\caption{
\label{fig:metal_ratios_b}
  The N(CIV)/N(CIV),
N(OVI)/N(CIV) and N(HeII)/N(HeI)
column density ratios vs.  N(CII)/N(CIV) as predicted
by the mini-halo model for the spectra shown in
Figure~\ref{fig:input_spectrum}.  The left panels plot the ratios
corresponding to the $\alpha = -1$ (solid line), $\alpha = -1.5$ w/break
(dotted line) and O star spectra (dashed line).  The right panels plot
the ratios from using the spectra generated by Haardt \& Madau 1996 at
three redshifts: z = 0 (solid line), z = 2 (dotted line) and z = 4
(dashed line). 
}
\end{figure}

  Figures~\ref{fig:metal_ratios_a} and \ref{fig:metal_ratios_b} show the
metal line column density ratios predicted by the mini-halo model for
all six spectra shown in Figure~\ref{fig:input_spectrum}.  Various ions
of C, O, Si and He are selected with transitions which can be detected
from the ground, with the exception of the He lines at $\lambda \ll$
912\AA, which must be observed from space.  (Note: See Table 4 in
\cite{Morton88} 1988 for transition wavelengths.  Oscillator strengths
for these lines are found in \cite{Morton91} 1991, and some important
recent f-value revisions are summarized by \cite{Tripp96} 1996.) The
ratio N(CII)/N(CIV) is chosen as the standard abscissa in this type of
plot because the CII and CIV lines are extremely strong and therefore
can be detected in low metallicity gas.  However, these lines are also
prone to saturation and must be used cautiously. 

\begin{figure}
\plotone{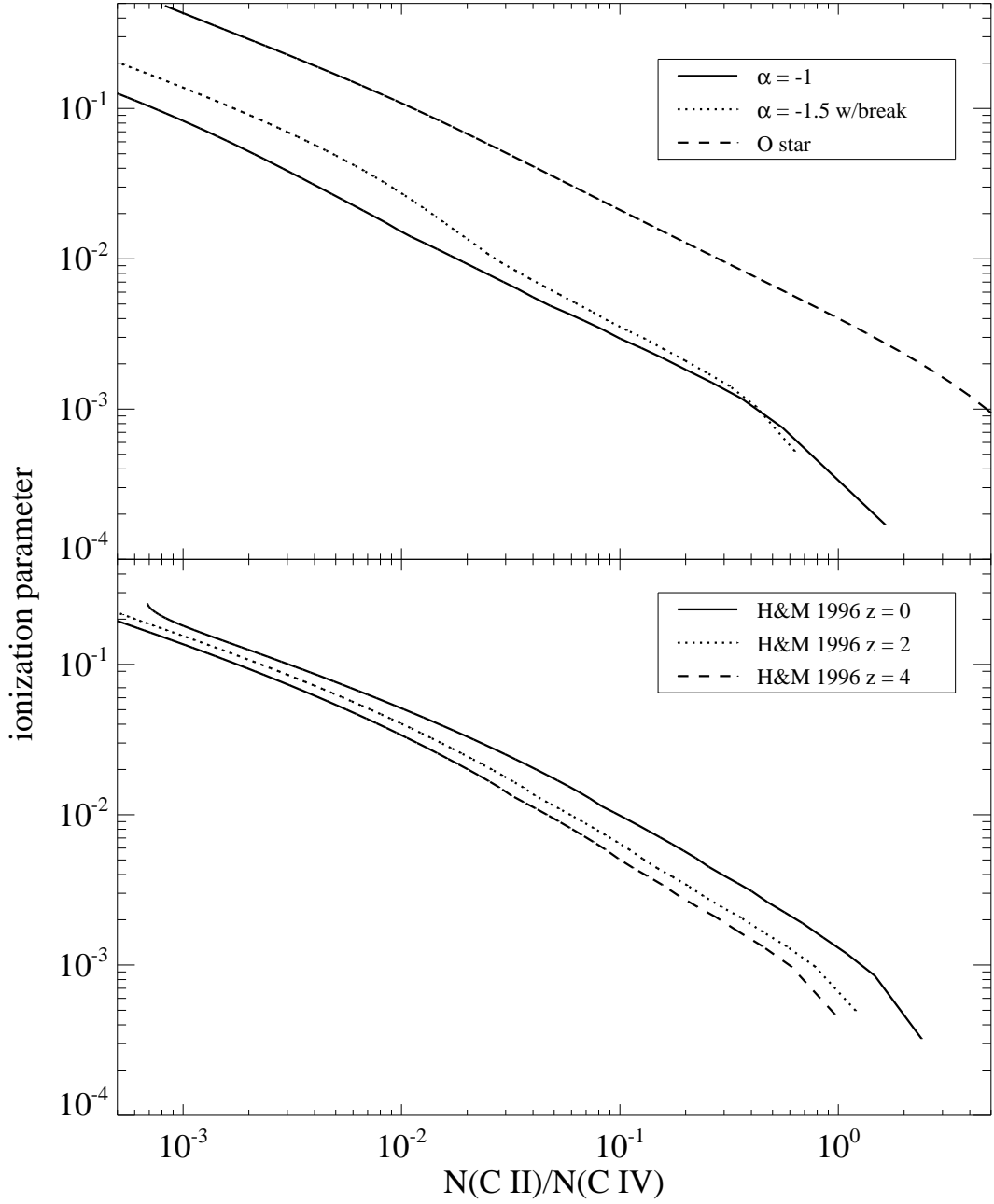}
\caption{
\label{fig:U_parameter}
  Ionization parameter, U, as function of N(CII)/N(CIV)
as computed from the mini-halo model for various input spectra.
}
\end{figure}

  The model ratios are primarily a function of the ionization parameter
U, the ratio of the H ionizing photon density to the total hydrogen
density, which decreases as N(CII)/N(CIV) increases.  The value of U for
the six input spectra is shown as a function of N(CII)/N(CIV) in
Figure~\ref{fig:U_parameter}.  In terms of the mini-halo model depicted
in Figures~\ref{fig:HI_den_evolution}-\ref{fig:HI_phase_col} low
N(CII)/N(CIV) corresponds to the outer part of the halo and high
N(CII)/N(CIV) corresponds to the inner part. 

  For the ratios shown in Figures~\ref{fig:metal_ratios_a} and
\ref{fig:metal_ratios_b}, U ranges from $10^{-1}$ to $10^{-4}$. 
Figure~\ref{fig:metal_ratios_a} shows that over this range of U, the
metal ratio differences resulting from the different input spectra (in
particular the $\alpha = -1.5$ with a HeII break and the \cite{Haardt96}
1996 spectra) are usually within a factor of 3, which is typically less
than the scatter in the observed ratios (see below).  Notable exceptions
are the O star ratios, which not surprisingly are quite different, even
from the $\alpha = -1.5$ w/break spectra, and the OVI/CIV and the
SiIV/CIV ratios which in some cases differ by two orders of magnitude. 

  The right panels of Figures~\ref{fig:metal_ratios_a} and
~\ref{fig:metal_ratios_b} show the mini-halo metal ratios at $z$ = 0, 2,
and 4 based on the \cite{Haardt96} (1996) modeling of the UV background, which
show very little change in the metal ratios between $z$ = 4 and $z$ = 2,
i.e., the abrupt change at $z$ = 3 in the SiIV/CIV ratio reported by
\cite{Songaila96} (1996) is not expected in mini-halos given the
\cite{Haardt96} (1996) UV background.  Again, the largest changes
(greater than a factor of 3) from $z$ = 4 to $z$ = 0 are in the OVI/CIV and
SiIV/CIV ratios; it seems that the other ratios do not change much over
this entire redshift range if the ionizing UV background is
predominantly due to QSOs and AGNs. 

\subsection{Metal Line Ratios in High Redshift Absorbers}

  In Figure~\ref{fig:metal_ratios_obs} observed column density ratios
from a sample of high redshift ($z > 2.161$) absorbers are plotted with
the mini-halo model ratios for three radiation fields: 
\cite{Haardt96} (1996) at $z = 2$, $\alpha =
-1.5$ w/break and O star.  The SiIV/CIV and SiII/CIV ratios are plotted
versus the CII/CIV ratio.  All of observed column densities in this
Figure were measured with the Keck 1 echelle spectrograph and include a
mix of Lyman limit and Ly$\alpha$ forest absorbers (shown with open
diamonds) from \cite{Songaila98} (1998) as well as four higher column density
damped Ly$\alpha$ absorbers (filled squares) from \cite{Lu96} (1996). 

  The damped absorbers are the four systems from the \cite{Lu96} (1996) sample
which have low enough HI column densities that the mini-halo model
still applies --- their column densities range from log N(HI) = 20.20
to log N(HI) = 20.52.  Most of these systems are complex
multiple-component absorbers, so a single mini-halo obviously cannot
produce the observed absorption profiles.  However, as noted in \S 1,
if these absorption systems are collections of merging smaller objects
as postulated in the hierarchical galaxy formation model, then the {\it
individual} clumps may be well-described by the mini-halo model. 
Alternatively, some of the absorption may be due to mini-halos
accreting onto bigger galaxies.  In either case, it would be useful
to compare these models to the observed properties of the individual
components, but unfortunately such data are not yet widely available. 
The large amount of data contained in Keck spectra makes it difficult
to provide a full listing of each component.  With a few exceptions
(e.g., \cite{Ganguly98} 1998), most Keck publications provide in tabular
format only the total column densities integrated over all components. 
Nevertheless, these total column densities provide weighted averages of
the column densities of the individual components and can be used to
test the validity of photoionization models. 

  Figure~\ref{fig:metal_ratios_obs} shows that the multi-phase mini-halo
model developed here leads to many of the same conclusions reached by,
among others, \cite{Giroux97} (1997) and \cite{Songaila98} (1998) on the
basis of single-phase constant density slab models.  Specifically, it
appears that the models with the straight power-law and the
\cite{Haardt96} (1996) ionizing radiation fields fall well below many of
the observed ratios when N(CII)/N(CIV) $\lesssim 10^{-1}$.  This problem
is alleviated by adopting the much softer O star spectrum.  The
power-law with a break at the HeII edge comes closer to the observed
ratios but still falls short.  Evidently in this model the flux must
drop by substantially more than a factor of 50 at the HeII break or Si
must be over-abundant relative to C (compared to the solar ratio) in
these high $z$ absorbers. 

\subsection{Helium Absorption Lines}

  Recently detected HeII absorption has raised the question as to
whether or not this absorption is entirely due to HeII associated with
the discrete HI Ly$\alpha$ clouds or whether a diffuse smoothly distributed 
IGM makes a contribution. A variety of calculations have been
performed to try to reproduce the observed HeII optical depth
(e.g., \cite{Giroux95} 1995; \cite{Fardal98} 1998).  These models are not 
completely constrained by the observations and require a variety of inputs 
(e.g. the amplitude, shape and evolution of the background radiation field). 
These authors find that a radiation field similar to \cite{Haardt96} (1996)
with N(HeII)/N(HI) $\sim$ 100 provides the best results.  The mini-halo
model is consistent with this value (see Figure~\ref{fig:HeII_HI}) for
absorbers with N(HI) $< 10^{16}$, which are believed to dominate the
discrete component of the HeII optical depth. 

  HeI absorption has also been claimed to be detected from space in four Lyman 
limit systems (\cite{Reimers92} 1992; \cite{Reimers93} 1993).  These results 
have been used to try and provide additional constraints on the the global He/H
value.  The limited number of observations make detailed comparisons
difficult, but the observed N(HeI)/N(HI) are consistent with the
mini-halo model if a harder ($\alpha = -1$) spectrum is used (see
Figure~\ref{fig:HeI_HI}), but this spectrum is unlikely to be consistent
with the metal line data.  Interestingly, similar slab based
calculations were unable to match these observations (\cite{Reimers93} 1993). 
These Lyman limit systems may correspond to a region along the core edge
where the multi-phase and geometric effects of the mini-halo model are
greatest (see Figures~\ref{fig:HI_phase_den} and
\ref{fig:HI_phase_col}).  If this were true, one might expect to see a
class of HeI absorbers corresponding to LL systems. However, we note that
the identification of the \ion{He}{1} absorption lines in the \cite{Reimers92} 
(1992) data is equivocal due to the low resolution of the data and requires 
verification with higher resolution spectroscopy.

\subsection{Low Redshift Absorbers}

  \cite{Barlow98} (1998) have recently shown that most low $z$ Ly$\alpha$
clouds do contain metals, and their metallicities are higher than the
high $z$ clouds.  STIS, the Cosmics Origins Spectrograph (COS - to be
installed in HST in 2003), and to a lesser extent FUSE will provide a
wealth of new data on metals in low $z$ absorbers.  As an example of the
type of analysis that will be possible, the mini-halo is compared with
the observations of the intervening absorber at z = 0.225 toward H
1821+643 (\cite{Savage98} 1998).  This absorber has a good measurement of the
OVI column density (log N(O VI) = 14.29$\pm$0.03) and upper limits on
CIV, SiIV and SiII (see Table 3 in \cite{Savage98} 1998). 
  %
  %
Figure~\ref{fig:OVI_absorber} shows these limits along with the
mini-halo model predictions.  Comparing the mini-halo OVI column density
in the allowed region of Figure~\ref{fig:OVI_absorber} for gas with 1/10
solar metallicity results gives a value of N(OVI) $\sim 10^{13}$
indicating that $ten$ mini-halos would be needed to explain this
absorber.  This is not allowed because the high resolution OVI
absorption profile shown in Figure 3 of \cite{Savage98} (1998) only shows
evidence of one component.  Furthermore, the good correspondence of the
red wing of the OVI and HI Lyman $\beta$ profiles in this absorber (again,
see Fig.  3 in \cite{Savage98} 1998) suggests that there is an appreciable
amount of HI absorption which occurs in the same gas that produces the
OVI absorption.  The mini-halo model will produce a negligible amount of
HI absorption if it is required to also satisfy the above constraints on
the metals.  Therefore the mini-halo model is unable to explain this
absorber on two counts (1) not enough OVI, and (2) not enough HI.  The
more likely explanation is a larger, more diffuse object. 

\section{Discussion}

  The previous sections presented a variety of results comparing
mini-halos with absorption line observations.  This section
considers the implication of a few of these results in more detail.  \S 5.1
examines the overall abundance of mini-halos.  In \S 5.2 looks at the
implications of the metal-line comparisons and some possible alternative
explanations.  \S 5.3 looks at other possible probes of the mini-halo
model.

\subsection{Abundance of Mini-Halos}

  The abundance of these mini-halos can be roughly calculated with
Press-Schechter formalism.  Although, it is important to keep in mind
that Press-Schechter theory is most accurate at the largest scales
(i.e., clusters of galaxies) and becomes more uncertain for estimating
the distribution of smaller objects.  Following the calculation in
\cite{Abel98} (1998), which uses the method of \cite{Lacey94} (1994) to
account for merging of halos, it is possible to estimate the number of
halos per unit redshift with $v_c \sim 30~\kms$ (see
Figure~\ref{fig:dNdz}).  These simple estimates assume column density
profiles similar to those calculated in the previous section.  Although
these calculations are far from definitive, they are consistent with the
observations of \cite{Storrie94} (1994) and \cite{Stengler95} (1995). 
It is not inconceivable that a significant fraction of absorption line
systems could be attributable to gas in mini-halos.  However, recent
observations and cosmological simulations suggest that it is unlikely
that all of the QSO absorption line systems are due to mini-halos (see
\S 1). 

 If some of the present day offspring of mini-halos are high velocity
clouds like those believed to be falling into the Milky Way
(\cite{Blitz98} 1998), then the abundance of these clouds gives an idea of
the likelihood of observing one of these objects near a large galaxy.
Estimates of the covering fraction of high velocity clouds around our
own galaxy would suggest a probability of finding a mini-halo within
1.5 Mpc of a larger galaxy is around 0.2. 

\subsection{Explaining the high redshift absorbers}

\begin{figure}
\plotone{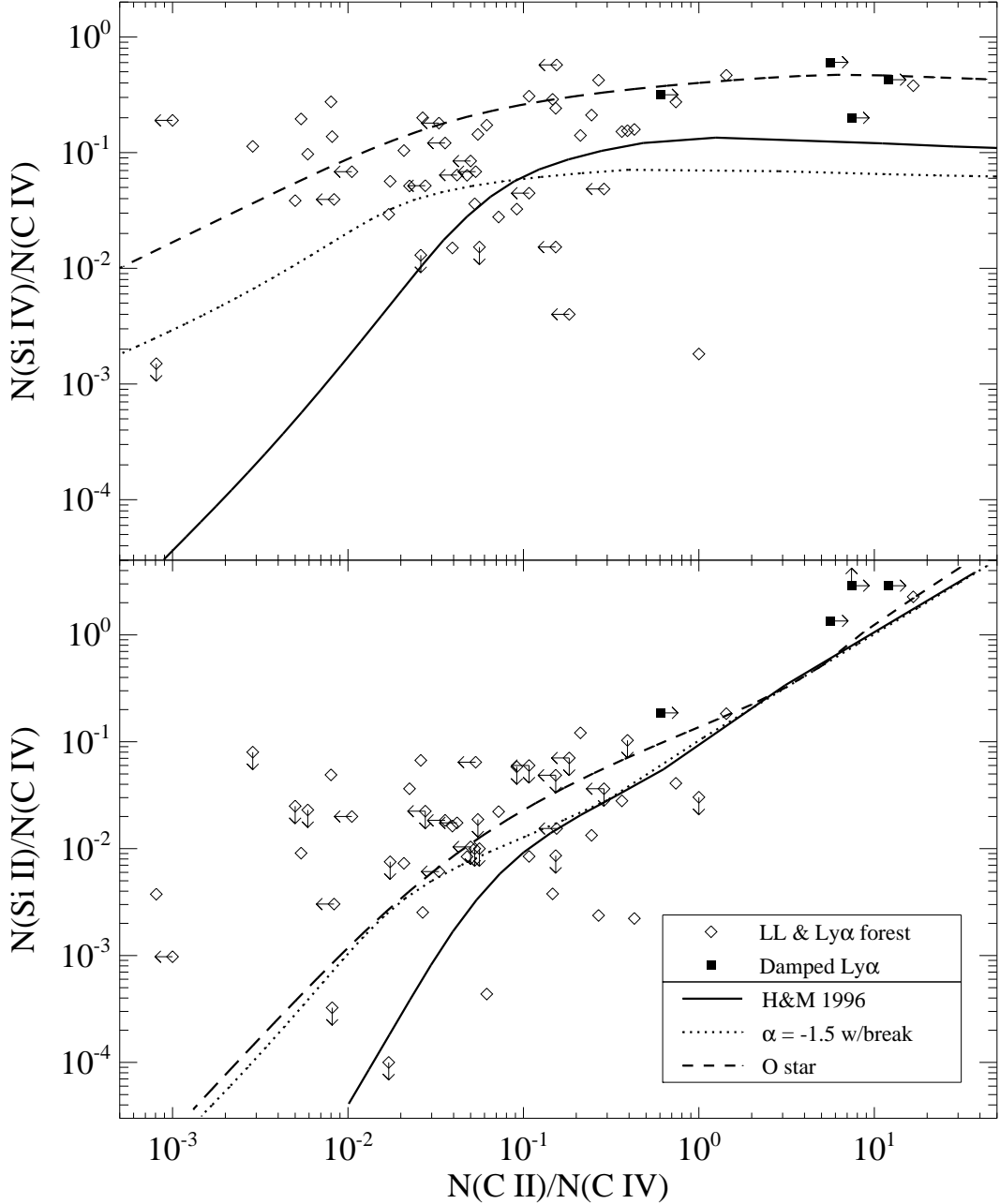}
\caption{
\label{fig:metal_ratios_obs}
  Observed column density ratios from a sample of high redshift ($z
>$ 2.1614) absorbers compared to the mini-halo model ratios for the
radiation fields: Haardt \& Madau 1996 $z = 2$ (solid line), $\alpha =
-1.5$ w/break (dotted line) and O star (dashed line).  The
N(SiIV)/N(CIV) (top panel) and N(SiII)/N(CIV) (bottom panel)
ratios are plotted versus the N(CII)/N(CIV) ratio.  The data
include a mix of Lyman limit and Ly$\alpha$ forest absorbers (diamonds)
(\cite{Songaila98} 1998) and four higher column density damped
Ly$\alpha$ absorbers (filled squares) (\cite{Lu96} 1996). 
}
\end{figure}

\begin{figure}
\plotone{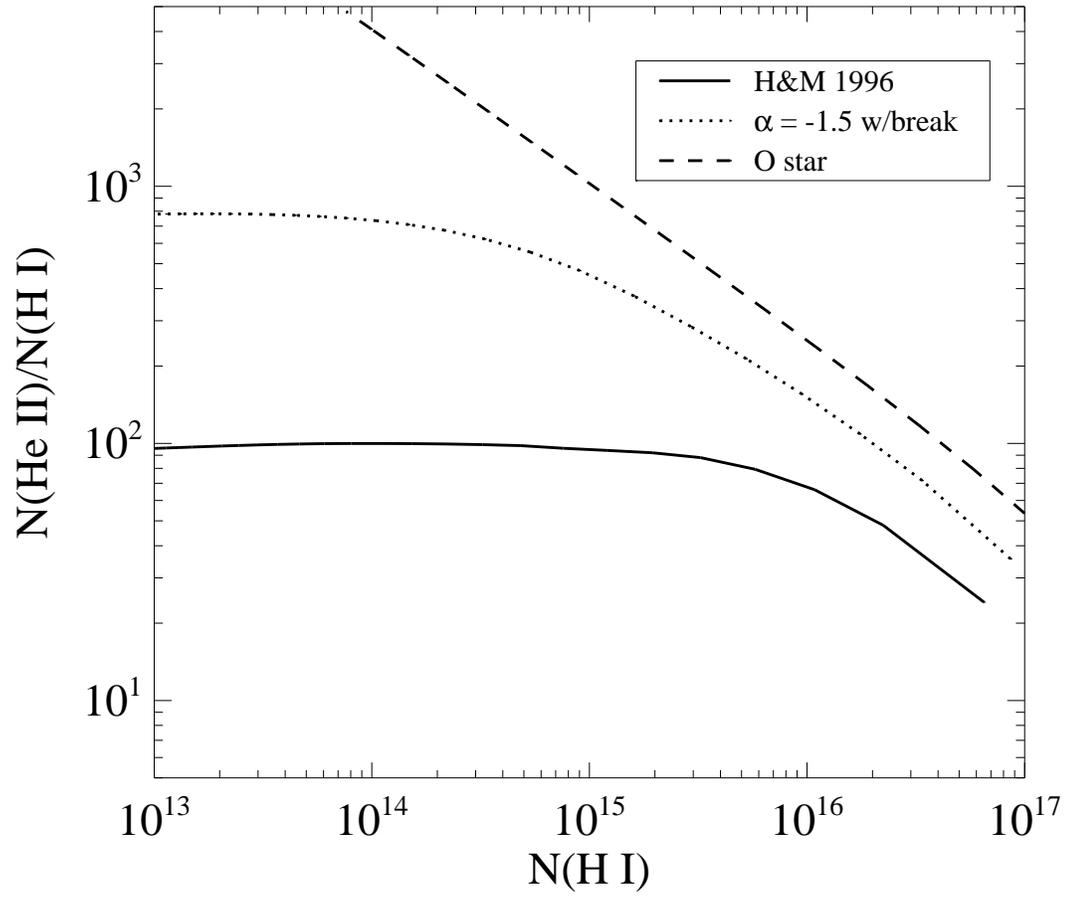}
\caption{
\label{fig:HeII_HI}
  The N(HeII)/N(HI) column density ratio vs. 
N(HI) as predicted by the mini-halo model for three radiation
fields: Haardt \& Madau 1996 $z = 2$ (solid line), $\alpha = -1.5$
w/break (dotted line) and O star (dashed line).
}
\end{figure}

\begin{figure}
\plotone{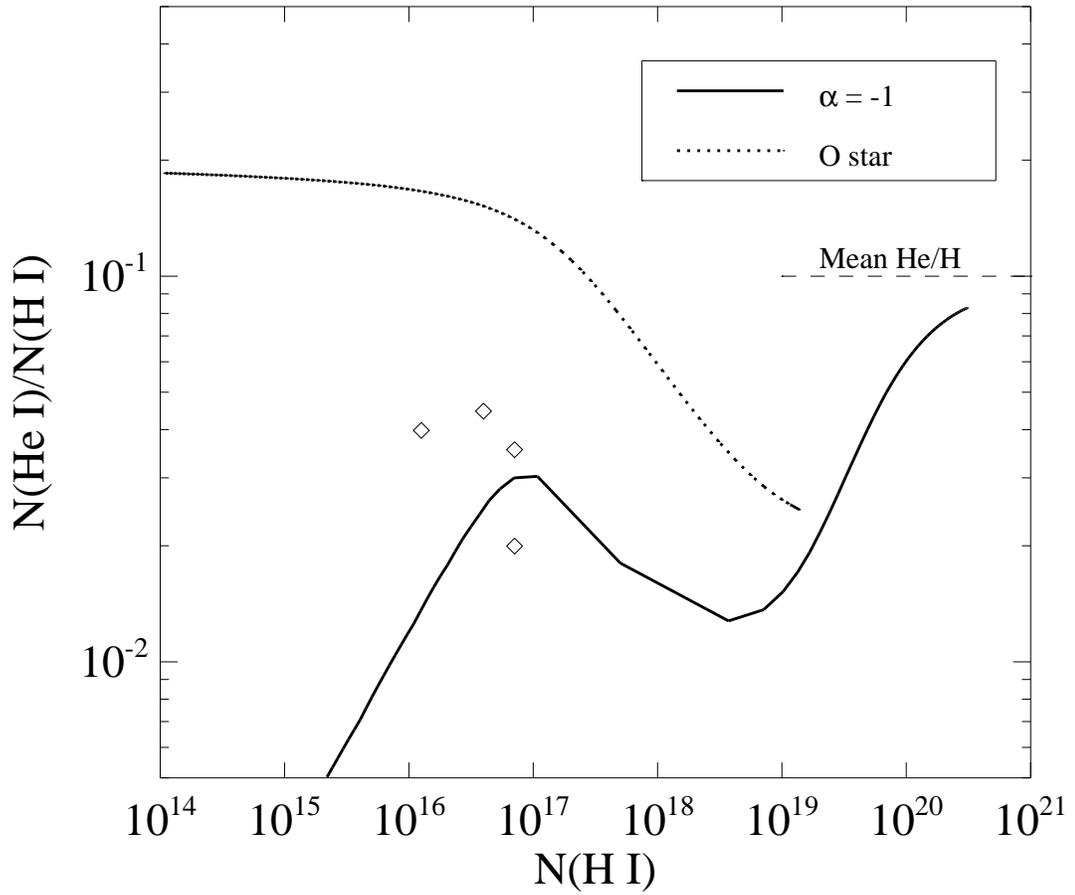}
\caption{
\label{fig:HeI_HI}
  The N(HeI)/N(HI) column density ratio vs.  N(HI) as predicted by the
mini-halo model for two radiation fields: $\alpha = -1$ (solid line) and
O star (dotted line).  The observations are taken from \cite{Reimers92} (1992) 
and \cite{Reimers93} (1993).
}
\end{figure}

\begin{figure}
\plotone{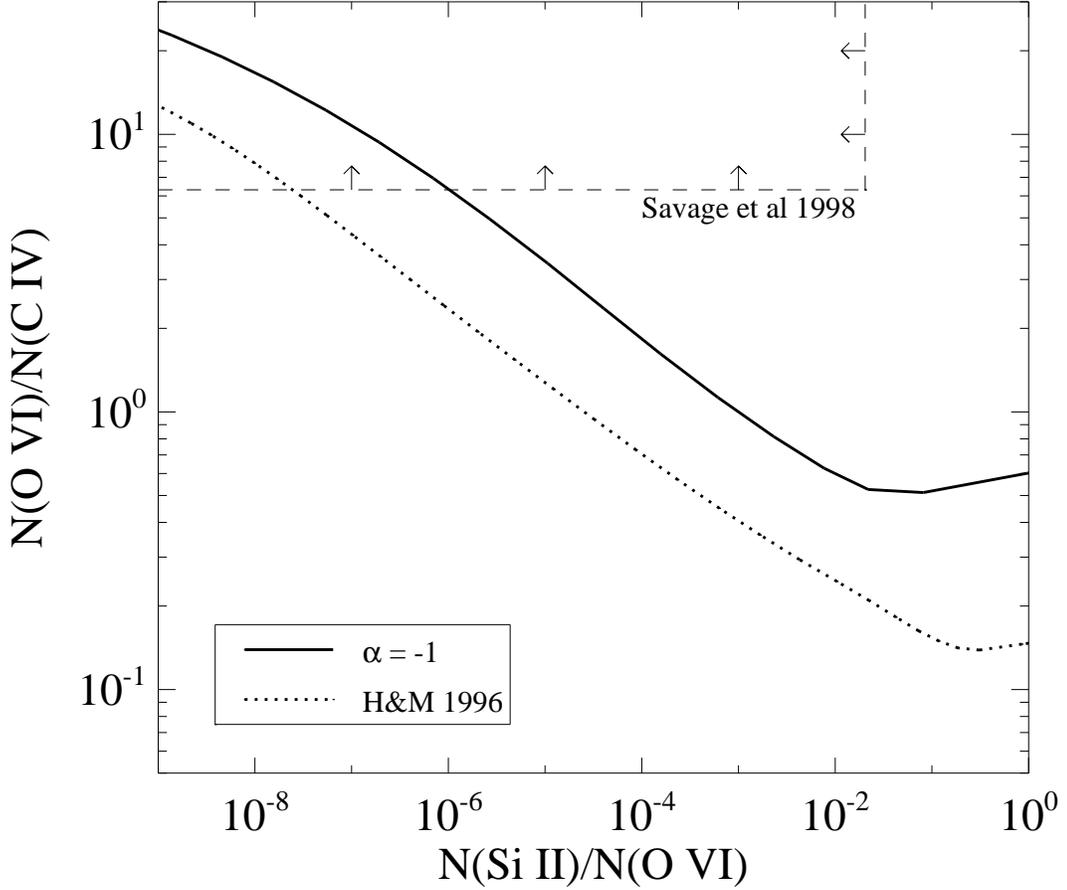}
\caption{
\label{fig:OVI_absorber}
  The N(SiII)/N(OVI), column density ratio vs. 
N(OVI)/N(CIV) as predicted by the mini-halo model for
the $\alpha = -1$ (solid line) and Haardt \& Madau 1996 z = 0 (dotted
line) spectra.  The box denoted by the dashed line are constraints taken
from the H 1821$+$643 absorber (see Table 3 in Savage et al.  1998): log
N(OVI) = 14.29$\pm$0.03, log N(CIV) $<$ 13.43,
log N(SiIV) $<$ 13.20, and log N(SiII) $<$ 12.54.
}
\end{figure}

\begin{figure}
\plotone{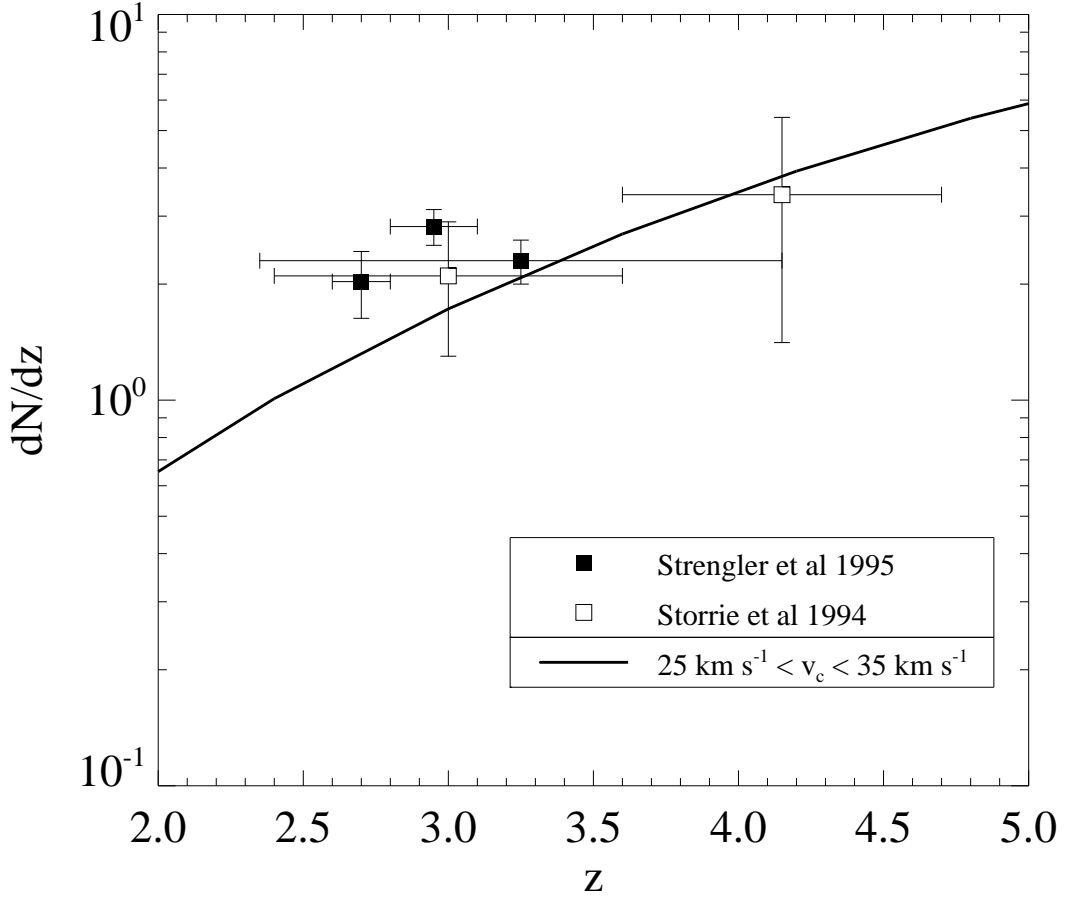}
\caption{
\label{fig:dNdz}
  Evolution of the number of dark matter halos per unit redshift in the
of halos with circular velocities in the range $25~\kms < v_c <
35~\kms$.  The estimates of dN/dz are based on Press-Schechter theory
with corrections for merging in an SCDM universe.  The points show the
number of observed Lyman limit systems. 
}
\end{figure}

  As shown in Figure~\ref{fig:metal_ratios_obs}, it appears that the
models with the straight power-law and the \cite{Haardt96} (1996) ionizing
radiation fields fall well below many of the observed ratios when
N(CII)/N(CIV) $\lesssim 10^{-1}$.  This problem is alleviated by
adopting the much softer O star spectrum.  The power-law with a break at
the HeII edge comes closer to the observed ratios but still falls short. 
If the flux increases at energies higher than the HeII break, as might
be expected due to the decreasing HeII absorption cross section with
increasing photon energy, then the observed ratios become even harder to
fit with the power-law + HeII break spectrum (\cite{Rauch97} 1997b).  The flux
may not recover as rapidly as expected due to smearing of the absorption
by He at different redshifts along the line of sight, but nevertheless
this scenario provides motivation for considering alternatives. 
\cite{Songaila98} (1998) suggests the power-law + HeII break ionizing spectrum
provides a natural explanation for the change in the SiIV/CIV ratio at
$z \approx 3$.  At $z > 3$ He is not yet fully reionized and
consequently the universe has substantial opacity at the He break.  At
$z < 3$, He reionization is complete so the He break vanishes leading to
smaller SiIV/CIV ratios. 

  An alternative is that the ionization of some of the absorbers is
dominated by radiation from hot stars (\cite{Giroux97} 1997).
Figure~\ref{fig:metal_ratios_obs} shows that if the ionizing radiation
can be approximately described by an O-star spectrum (dashed line in
Fig~\ref{fig:metal_ratios_obs}), then the observed ratios are reproduced
by the model.  In this respect, it is interesting to note that the
highest SiIV/CIV ratios measured by \cite{Songaila98} (1998) are quite similar
to the SiIV/CIV ratios observed in the ISM of the Milky Way.  For
example, \cite{Sembach97} (1997) have measured the CIV/SiIV ratio in the
Galactic ISM using sight lines to 31 stars observed with {\it HST} and
{\it IUE}, and they derive N(CIV)/N(SiIV) = 3.8$\pm$1.9, close to the
highest ratios in Figure~\ref{fig:metal_ratios_obs}.  The high
ionization of these Milky Way sight lines is probably produced by
photoionization from hot stars or collisional ionization with additional
photoionization from re-emission in the hot gas (see \S 8 in
\cite{Sembach97} 1997).

  It is not entirely unreasonable to suggest that hot stars make a
substantial contribution to the photoionization of high $z$ absorbers. 
At some point in the past, galaxies must have undergone widespread waves
of star formation.  Based on the apparent increase in the metallicity of
damped Ly$\alpha$ absorbers at $z \approx 3$, \cite{Lu96} (1996) have suggested
that this redshift marks the first epoch of major star formation in
galaxies.  This star formation will be accompanied by copious UV
emission, and if this UV light is able to escape, then it could be a
substantial source of photoionization. 

  In this scenario, what are some possible explanations for the change
in the SiIV/CIV ratio at $z$ = 3 reported by \cite{Songaila98} (1998)? One
possibility is that the SiIV/CIV ratio is a function of absorber HI
column density, and the portion of Songaila's sample with $z >$ 3 has a
larger fraction of systems with higher HI columns.  In this case, the
different \ion{Si}{4}/\ion{C}{4} ratios could be caused
by self-shielding due to He and/or H in higher column density absorbers,
for example.  The \cite{Songaila98} (1998) sample contains Lyman limit 
[N(HI) $\gtrsim 10^{17}$ cm$^{-2}$] and Ly$\alpha$ forest absorbers, but
\ion{H}{1} column densities are not provided.  However, the CII and SiII 
column densities can be used as rough proxies for the HI column since 
absorbers with higher N(HI) will have higher CII and SiII column densities 
as well. 

\begin{figure}
\plotone{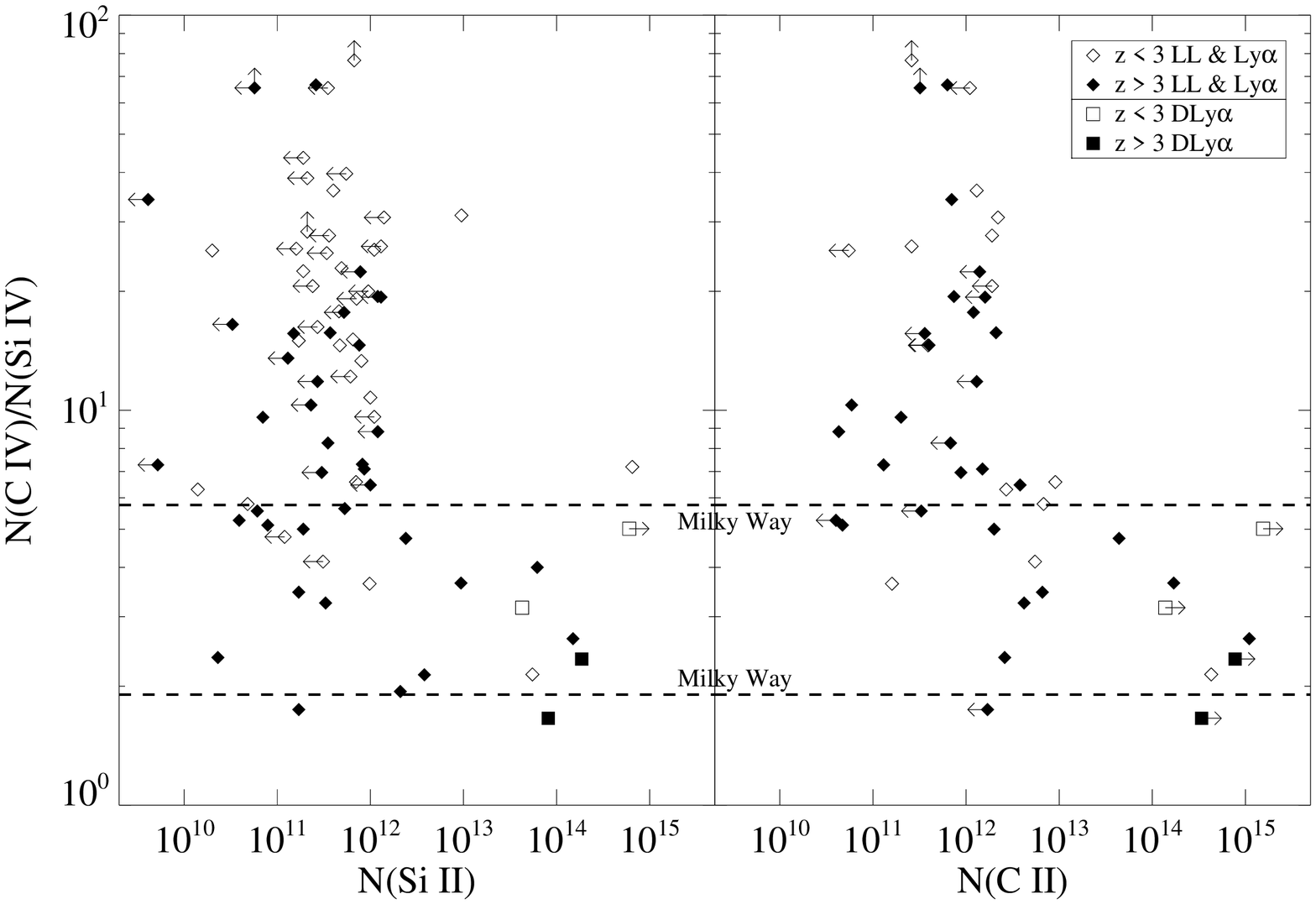}
\caption{
\label{fig:metal_ratios_z_dep}
  Observed column density ratios from a sample of high redshift ($z
>$ 2.1614) absorbers.  N(CIV)/N(SiIV) ratio is plotted
vs. N(SiII) (left panel) and N(SiII) (right panel). 
The observed column densities in this Figure include a mix of Lyman
limit and Ly$\alpha$ forest absorbers (diamonds, \cite{Songaila98} 1998) as
well as four higher column density damped Ly$\alpha$ absorbers (squares)
from \cite{Lu96} (1996).  Open symbols correspond to absorbers with $z < 3$
while filled symbols indicate $z > 3$.  The ratio computed for the Milky
Way is shown between the two dashed lines. 
}
\end{figure}

  To check for a possible N(HI) dependence,
Figure~\ref{fig:metal_ratios_z_dep} plots the CIV/SiIV column density
ratio from the \cite{Songaila98} (1998) sample (diamond symbols) vs.  N(SiII)
(left panel) and vs.  N(CII) (right panel).  [Note: the SiII
detections or limits are available for most of the absorbers in the
\cite{Songaila98} (1998) sample while CII measurements or limits are available
for a fraction of the systems.] In addition, the four damped systems
from \cite{Lu96} (1996, square symbols) have also been added.  In this Figure,
$z > 3$ absorbers are indicated with filled symbols while $z < 3$
systems are marked with open symbols, and systems with only upper or
lower limits are shown with arrows.  Finally the Milky Way value of CIV/SIV
obtained by \cite{Sembach97} (1997) is bracketed by the horizontal dashed lines. 

  Figure~\ref{fig:metal_ratios_obs} suggests several interesting points:
\BENUM
  \item As mentioned earlier, there appears to be a
  significantly larger fraction of systems with low N(CIV)/N(SiIV) at $z > 3$.
  \item Many of the high column systems have N(CIV)/N(SiIV)
        ratios consistent with the ratios in our own Milky Way ISM.
   \item There are almost no high column systems (N(SiII) $> 2 \times
   10^{12}$ or N(CII) $> 6 \times 10^{12}$) with high N(CIV)/N(SiIV) at any
   redshift. 
   \item There appears to be a significantly larger fraction of high column
   systems (N(SiII) $> 2 \times 10^{12}$ or N(CII) $> 6 \times 10^{12}$) at
   $z > 3$. 
\EENUM
One hypothesis that is consistent with these points (the third and the
fourth in particular) is that the lower column systems with higher
N(CIV)/N(SiIV) exist but were not detected at $z > 3$.  However, for
this hypothesis to be true, there must also be a dependence of
N(CIV)/N(SiIV) on column density.  Thus, it would be useful to examine
the dependence of the CIV/SiIV ratio on HI column density using direct
measurements of N(HI). 

\subsection{Other probes}

  The absorption signatures discussed so far are for species whose
populations are dominated by the background radiation field.  To fully
probe the multi-phase density and temperature structure of the mini-halo
requires examining other species that are involved in more complex
chemistry, such as molecular hydrogen.  So far H$_2$ detections are
rare, but as they become more common the mini-halo model can be used to
make definite predictions.  These may be particularly useful in higher
column systems that correspond to the core of the mini-halo where H$_2$
is most prevalent (see Figures~\ref{fig:HI_phase_den} and
\ref{fig:HI_phase_col}).

  Another useful probe might be CII*.  CII is a strong line that is often
detected and CII* has been detected in a few cases.  For illustration
purposes,the value of CII* is calculated assuming statistical balance
between collisional de-excitation and collisional + radiative excitation
between the level 0 ($^2$P$_{\frac{1}{2}}$) and level 1 ($^2$P$_{\frac{3}{2}}$)
states:
\BEQA
    \frac{n({\rm CII*})}{n{\rm (CII)}} & = &
       \frac{C_{01}}{A_{10} + C_{10} + C_{01}}, \NN \\
     C_{10} & = & 1.81 \times 10^{-5} T^{-0.5} n(e^-) + \NN \\
            &   & 5.0 \times 10^{-10} [1 + 0.12 T^{0.5}] n({\rm HI}) + \NN \\
            &   & 4.86 \times 10^{-10} (0.01 T)^{0.11} [n({\rm H_2}) + 0.5 n({\rm HeI})], \NN 
\\
     C_{01} & = & 2 \exp[-91.211/T] C_{10}, \NN \\
     A_{10} & = & 2.4 \times 10^{-6},
\EEQA
where the coefficients are computed as follows: HI (\cite{Harel78} 1978),
H$_2$ (\cite{Flower77} 1977), He (assumed to be 0.5 times H$_2$), e$^-$
(\cite{Keenan86} 1986) and A$_{10}$ (\cite{Nussbaumer88} 1988).  At maximum
N(CII*)/N(CII) $\sim 10^{-5}$, which is unlikely to be detected and
more importantly is swamped by CII* excitation by the
cosmic microwave background (\cite{Lu96} 1996).  In any case, additional
diagnostics like these can be explored and tested with the mini-halo
model.

\section{Conclusions and Further Work}

  In this paper the absorption line signatures of gas in mini dark
matter halos were computed.  The motivation for this comes from two
perspectives.  First, mini-halos may exist in significant numbers and be
the source of many absorption line systems.  Second, absorption line
systems in general may be multi-phase objects and the mini-halo model
provides one means of exploring multi-phase structures. 

  The mini-halo model can be compared with a wide variety of
observations both ground and space based.  A few such comparisons are
made here: high redshift metal line systems, He absorption line systems
and a low redshift OVI absorber.  It appears that the mini-halo model is
consistent with the high redshift metal line systems and leads to many
of the same conclusions made with single-phase slab models.  In
particular, a very soft spectrum is required to fit the observed
N(SiIV)/N(CIV) data.  Likewise, the mini-halo model N(HeII)/N(HI) is
consistent with calculations by other researchers based on the observed
HeII optical depth.  Perhaps, one interesting result is that the
mini-halo model with a $\alpha = -1$ spectrum is consistent with the HeI
observations in Lyman limit systems, where other models have had a more
difficult time matching this result.  Finally, a comparison with a
single OVI absorber indicates that the mini-halo is not a good
explanation for this object and that a larger more diffuse system is
more likely. 

  Based on these results it appears that the mini-halo model absorption
line signatures are consistent with most current observations.  Thus, in
any given instance it is difficult to either rule in or rule out a
mini-halo and in most cases additional data (e.g.  optical counterparts
or the lack thereof) or a detailed examination of predicted line
profiles may be necessary to break the degeneracy.  There are a number
of simplifications in the model presented here, and several issues
should be explored in future studies.  First, the effects of any star
formation on the mini-halos should be considered; even a very small
number of supernovae could substantially affect the structure of a
mini-halo by driving a wind and blowing out gas.  This may increase the
likelihood that mini-halos will cause QSO absorption lines by increasing
their spatial cross sections.  Second, the impact of shock heating on
the mini-halo absorption line signatures should be evaluated.  Shocks
are generated by the collapse of initial density perturbations, and
these are seen to cause substantial heating of gas in the vicinity of
galaxies in cosmological hydrodynamic simulations (e.g., \cite{Dave99}
1999; \cite{Cen99} 1999).  Finally, the input radiation fields can be
refined.  For example, a considerably more sophisticated model for the
radiation from a star forming galaxy than the simple O-star spectrum
should be developed.  New hot star model atmospheres emit more ionizing
continuum (e.g., \cite{Schaerer97} 1997), and this may alter the metal
line ratios.  The mini-halo model is a useful tool for analyzing the
absorption line data in a multi-phase context and should become even
more applicable as new space based observations become available. 

\acknowledgments

We would like to thank Bruce Draine for his assistance with the CII*
calculations.  We would also like to thank Ed Fitzpatrick for providing
the O star spectrum and Francesco Haardt for making available the
\cite{Haardt96} spectra.  Jeremy Kepner was supported by NSF grant
AST~93-15368.  Tom Abel was supported in part by NASA grant NAG5-3923.


\newpage


\singlespace

\end{document}